\documentclass[12pt,letter]{article}
\usepackage{amsfonts}
\usepackage{amssymb}
\usepackage{amssymb}
\usepackage{amsmath, setspace}
\usepackage{amsmath, amssymb, tikz, tabularx, pgfplotstable}
\usepackage{pgfplots,tikz, verbatim}
\usepackage{fixltx2e}
\usepackage{siunitx}
\usepackage{xcolor}
\usepackage[version = 4]{mhchem}
\usepackage{pgfplots}
\usepackage{a4}
\usepackage[a4paper,left = 3cm,right = 3cm,top = 3.1cm,bottom = 3.3cm]{geometry}

\newcommand{\MaxBusinessCycle}{100}
\newcommand{\MinBusinessCycle}{-100}
\usetikzlibrary{patterns}
\usepgfplotslibrary{external}
\pgfplotsset{compat = newest} 
\usepgfplotslibrary{groupplots}
\usetikzlibrary{matrix,calc}
\DeclareSIUnit{\molar}{M}
\pgfplotsset{compat = newest}
\usepgfplotslibrary{fillbetween}

\usepgfplotslibrary{groupplots}
\usetikzlibrary{pgfplots.groupplots}
\usetikzlibrary{plotmarks}
\usetikzlibrary{patterns}
\usepgfplotslibrary{external}
\pgfplotsset{compat = newest} 

\input{texdraw}
\usepgfplotslibrary{external}
\usepgfplotslibrary{groupplots}

\newcounter{lemma_counter}

\newtheorem{assumption}{Assumption}

\newtheorem{lemma}[lemma_counter]{Lemma}

\renewcommand{\Pr}{\mathbb{P}}

\providecommand{\keywords}[1]{\noindent \textbf{Keywords:} #1 \\}
\providecommand{\jelcodes}[1]{\noindent \textbf{JEL-codes:} #1 \\}

\renewcommand{\Pr}{\mathbb{P}}

\renewcommand{\mid}{\,|\,}

\providecommand{\keywords}[1]{\noindent \textbf{Keywords:} #1 \\}
\providecommand{\jelcodes}[1]{\noindent \textbf{JEL-codes:} #1 \\}
\begin{document}

\title{\textbf{Selection and the Distribution of Female Hourly
Wages in the U.S.}\thanks{
\ We are grateful to useful comments and suggestions from the editor Chris
Taber, two anonymous referees, St\'{e}phane Bonhomme, Chinhui Juhn, Ivana
Komunjer, Yona Rubinstein, Sami Stouli, and Finis Welch.}}
\author{Iv\'{a}n Fern\'{a}ndez-Val\thanks{%
\ Boston University } \and Aico van Vuuren\thanks{%
\ Gothenburg University and University of Groningen.} \and Francis Vella%
\thanks{%
\ Georgetown University } \and Franco Peracchi\thanks{%
University of Rome Tor Vergata}}
\maketitle

\begin{abstract}
We analyze the role of selection bias in generating the changes in the
observed distribution of female hourly wages in the United States using CPS
data for the years 1975 to 2020. We account for the selection bias from the
employment decision by modeling the distribution of the number of working
hours and estimating a nonseparable model of wages. We decompose changes in
the wage distribution into composition, structural and selection effects.
Composition effects have increased wages at all quantiles while the impact
of the structural effects varies by time period and quantile. Changes in the
role of selection only appear at the lower quantiles of the wage
distribution. The evidence suggests that there is positive selection in the
1970s which diminishes until the later 1990s. This reduces wages at lower
quantiles and increases wage inequality. Post 2000 there appears to be an
increase in positive sorting which reduces the selection effects on wage
inequality.
\end{abstract}

\keywords{Wage inequality, wage decompositions, nonseparability, selection
bias}

\vspace{-0.9cm} \jelcodes {C14, I24, J00}

\thispagestyle {empty}

\newpage \setcounter {page}{1}

\section{Introduction}


The dramatic increase in female wage inequality in the United States since
the early 1980s (see, for example, Katz and Murphy 1992, Katz and Autor
1999, Lee 1999, Autor et al.\ 2008, Acemoglu and Autor 2011, Autor et al.\
2016 and Murphy and Topel 2016) has been accompanied by substantial changes
in both female employment rates and the distribution of their annual hours
of work. Given the prominence that accounting for the selection bias (see
Heckman 1974, 1979) from employment decisions has played in empirical
studies of the determinants of female wages it seems natural to investigate
its role in the evolution of wage inequality. This paper examines the
sources of changes in the distribution of female hourly real wage rates in
the United States from 1975 to 2020 while accounting for movements, and
individuals' locations, in the annual hours of work distribution.

The inequality literature allocates wage changes to two sources. The first
is the \textquotedblleft structural effect" which captures the market value
of an individual's characteristics. This includes skill premia, such as the
returns to education (see, for example, Welch 2000 and Murphy and Topel
2016), cognitive and noncognitive skills (Heckman, Stixrud and Urzua 2006),
declining minimum wages in real terms (see, for example, DiNardo et al.\
1996 and Lee 1999), and the increasing use of non-compete clauses in
employment contracts (Krueger and Posner 2018). The second source, referred
to as the \textquotedblleft composition effect", reflects differences across
workers' observed characteristics. These include increases in educational
attainment. Earlier papers (see, for example, Angrist et al.\ 2006, and
Chernozhukov et al.\ 2013) have estimated these two effects under general
conditions. However, as they focus on male wages they have ignored the role
of selection.

Understanding the role of selection in the female wage inequality context is
important. First, as the impact of selection is frequently interpreted as
reflecting sorting patterns it is valuable from a policy perspective to
understand how worker productivity has changed as an increasing proportion
of women have entered the labor market. Second, the importance of accounting
for selection bias in estimating the determinants of female wages suggests
that an evaluation of the role of structural and composition effects
requires an appropriate treatment of selection. Third, assessing the impact
of selection on wages and inequality is particularly relevant when the
composition of the working and nonworking populations have evolved as
drastically as has occurred in our sample period. Finally, understanding the
impact of selection bias may provide policy makers with guidance as to which
measures may be taken to reduce wage inequality.

Three important papers have investigated the role of selection in the United
States over a period of increasing female wage inequality.\footnote{%
Fern\'{a}ndez-Val et al. (2022) study the impact of annual hours worked on
inequality but their focus is on annual earnings and not wages.} Mulligan
and Rubinstein (2008), hereafter MR, correct for selection in the female
mean wage in the United States for the years 1975-1999 and argue that the
sharp increase of female wages partially reflected that the selected
population of working females became increasingly more productive in terms
of unobservables. They also find the pattern of sorting turned from negative
to positive in the early 1990s. Evaluating the contribution of selection
bias on the mean wage is straightforward in additive models as the selection
component, under some assumptions, can be separated. The pattern of sorting
is inferred from the coefficient for the selection correction. The
nonseparable model required for estimating wage distributions has greater
difficulties in isolating the selection component. Maasoumi and Wang (2019),
hereafter MW, employ the copula based estimator for quantile selection
models of Arellano and Bonhomme (2017), hereafter AB. AB and MW define the
selection effect as the difference between the observed wage distribution
and the counterfactual wage distribution simulated via their models'
estimates assuming 100\% participation. MW provide a similar conclusion
regarding the pattern of sorting as MR for the overlapping years in their
studies. Blau et al. (2021) follow Olivetti and Petrongolo (2008) who use a
``selection on unobservables" approach to impute wages for nonworkers based
on propensity scores for employment. They also compute the predicted wage
distribution assuming 100\% participation. In contrast to MR and MW, they
find a more modest role for selection and that sorting did not change sign
over the sample period. We define the selection effect as the difference in
the observed wage distribution and the wage distribution that would result
under the participation process associated with the year with the lowest
participation rate. We find that the direction of sorting did not change
during the years considered by MR.

We address several methodological and empirical issues regarding selection
in the female wage inequality context. Our methodological contributions are
the following. First, we extend the Fern\'{a}ndez-Val, Van Vuuren, and Vella
(2021), hereafter FVV, estimator for nonseparable models with censored
selection rules. The FVV estimator incorporates the number of working hours
rather than the binary work decision as the selection variable and here we
allow for different censoring points conditional on the individual's
characteristics. This variation across censoring points captures differences
in \textquotedblleft fixed working costs" (see, for example, Cogan 1981).
Second, we provide a procedure for decomposing changes in the wage
distribution into structural, composition and selection effects in a
nonseparable model which allows for selection from the choice of annual
working hours. We contrast our decomposition approach with the corresponding
exercise based on the Heckman (1979) selection model (HSM). Third, we extend
our estimator to allow selection into annual hours to reflect two separate
selection mechanisms. Namely, the choices of annual weeks and weekly hours.
Fourth, we provide an estimator motivated by the ordered treatment model of
Heckman and Vytlacil (2007) which allows for bunching in annual hours or
annual weeks and apply it via our decomposition method.

Our following empirical contributions feature results based on the two most
commonly employed Current Population Survey (CPS) data sets. First, unlike
MR and MW who analyze wages for full-time full-year (FTFY) workers, we
obtain a fuller picture of the evolution of the wage distribution by
including all workers and accounting for selection from the hours of work
decision. Second, we confirm previous findings, restricted to FTFY workers,
regarding movements in the wage distribution. Female wage growth at lower
quantiles is modest although the median wage has grown steadily. Gains at
the upper quantiles are large and have produced an increase in female wage
inequality. Finally, we provide new evidence regarding the role of
selection. \ Changes in selection are especially important at the lower end
of the wage distribution and have generally decreased wage growth and
increased wage inequality. Although we are able to reproduce the estimated
sorting pattern as MR and MW, we illustrate this reflects the employed
identification assumptions. We show that exploiting the variation in hours
worked as a form of identification produces results consistent with positive
sorting for the whole sample period.

An important empirical result relates to the pattern of sorting and its
implication for the impact of selection on wages and inequality. We find
clear evidence of positive sorting in the mid 1970s. The period 1975 to 2000
experiences a shift in the distribution of female annual hours of work,
accompanied by a reduction in the level of positive sorting. These two
forces decrease wages at lower quantiles and increase wage inequality. For
the remainder of our sample period there appears to be a return to higher
levels of positive sorting and a decrease in the impact of selection on wage
inequality.

The rest of the paper is organized as follows. The next section discusses
the data. Section~\ref{model} describes our empirical model and defines our
decomposition exercise. It also provides alternative estimators employing
ordered or multiple censored selection rules. This section concludes with a
comparison of our decomposition approach with that associated with the HSM.
Section~\ref{results} presents the empirical results. Section \ref%
{sec:comparison} reconciles the difference between our results with those of
MR, while Section \ref{sec:inequality} investigates the impact of the
changes in selection in wage inequality. Section \ref{conclusions} offers
some concluding comments.


\section{Data}

\label{data} 

We employ the two most commonly analyzed micro-level data sets, the Annual
Social and Economic Supplement (ASEC) and the Merged Outgoing Rotation
Groups (MORG), from the CPS. Appendix A of Lemieux (2006) provides a
comparison of the two data sets. We employ both to contrast results and to
allow comparisons with earlier studies.

\subsection{Annual Social and Economic Supplement}

We employ the ASEC for the 46~survey years from 1976 to 2021 reporting
annual earnings for the previous calendar year.\footnote{%
The data were obtained via the IPUMS-CPS website maintained by the Minnesota
Population Center (Flood et al.\ 2015).} Unless otherwise stated, we refer
to the year for which the data are collected and not that of the survey. The
1976 survey is the first for which information on weeks worked and usual
hours of work per week last year are available. To avoid issues related to
retirement and ongoing educational investment we restrict attention to those
aged 24--65 years in the survey year. This produces an overall sample of
2,219,820 females. The annual sample sizes range from 33,924 in 1976 to
59,622 in 2001.

Annual hours worked are defined as the product of weeks worked and usual
weekly hours of work last year. Those reporting zero hours usually respond
that they are not in the labor force in the week of the March survey. We
define hourly wages as the ratio of reported annual labor earnings in the
year before the survey, converted to constant 2019 prices using the consumer
price index for all urban consumers, and annual hours worked. Hourly wages
are unavailable for those not in the labor force. For the self-employed,
unpaid family workers and the Armed Forces annual earnings or annual hours
tend to be poorly measured and we exclude these groups from our sample. This
results in a deletion of 5.4, 0.4 and 0.07 percent of the observations for
the self employed, unpaid family workers and the Armed Forces, respectively.
The figures for self employed and the armed forces have trended upwards
while those for family workers have trended downwards over the sample
period. These groups do not show any cyclical variation. The only exception
is the number of self employed during the Great Recession which, compared to
the total employed, dropped considerably.\label{insert:self_employed_out} We
use observations with imputed wages for their values of working hours but do
not use them in the wage sample.\label{insert:imputed} The restriction to
civilian dependent employees with positive hourly wages and people out of
the labor force last year results in a sample of 2,055,063 females. The
subsample of civilian dependent employees with positive hourly wages
comprises 1,190,928 observations. A benefit of the ASEC is its extensive
family background variables.

\subsection{Merged Outgoing Rotation Groups}

We use the years 1979 to 2019 for the MORG using the CEPR extracts. The MORG
contains information on hourly wages in the survey week for those paid by
the hour and on weekly earnings from the primary job during the survey week
for those not paid by the hour. Lemieux (2006) and Autor et al.\ (2008)
argue that the ORG CPS data are preferable as they provide a point-in-time
wage measure and workers paid by the hour, more than half of the U.S.\
workforce, may better recall their hourly wages. Based on comparable data
restrictions as the ASEC, the MORG results in 4,298,682 observations with
the highest numbers of observations in 1980 (121,786) and the lowest in 2019
(91,647). The subsample of civilian dependent employees working in the
reference week is 2,219,820 observations. This low figure, relative to the
ASEC, is expected as employees who did not work in the reference week may
have worked in another week. Family background variables are only available
since 1984. This restricts the family background characteristics to family
size.

\subsection{Descriptive statistics}

\begin{figure}[tbp]
\centering
\begin{tikzpicture}
	\begin{groupplot}
	[
	group style = {%
		group size = 3 by 2,
		group name = plots, 
		horizontal sep = 1cm, 
		vertical sep = 2cm,
		xlabels at = edge bottom,
		ylabels at = edge left
	},
	set layers,cell picture = true,
	width = 0.35\textwidth,
	height = 0.35\textwidth,
	legend columns = 1s,
	xlabel = Year,
	ylabel = Wage in dollars,
	xmin = 1975, 
	xmax = 2020,
	cycle list name = black white
	]
	\nextgroupplot[legend to name = grouplegend1, title = D1, ymin = 5, ymax = 12]
	\draw[dashed] (axis cs:1975,0)--(2020,0);
	\addplot[very thick] table [x expr={\thisrow{year}-1}, y=d1]{"quantiles.txt"};
	\addlegendentry{ASEC}
	\addplot[very thick, dashed] table [x =year, y=d1]{"quantiles_MORG.txt"};
	\addlegendentry{MORG}
	\addplot[domain=1981.0:1981.6, name path = C]{100};
	\addplot[domain=1981.0:1981.6, name path = D]{-1};      
	\addplot[domain=1982.6:1983.92, name path = E]{100};
	\addplot[domain=1982.6:1983.92, name path = F]{-1};      		
	\addplot[domain=1991.6:1992.25, name path = G]{100};
	\addplot[domain=1991.6:1992.25, name path = H]{-1};   	 
	\addplot[domain=2002.25:2002.92, name path = I]{100};
	\addplot[domain=2002.25:2002.92, name path = J]{-1};   			
	\addplot[domain=2008.09:2010.5, name path = K]{100};
	\addplot[domain=2008.09:2010.5, name path = L]{-1};   		
	\addplot[domain=2020.17:2020.33, name path = M]{100};
	\addplot[domain=2020.17:2020.33, name path = N]{-1};   	
	\addplot[lightgray] fill between[of=C and D];
	\addplot[lightgray] fill between[of=E and F];
	\addplot[lightgray] fill between[of=G and H];           
	\addplot[lightgray] fill between[of=I and J];
	\addplot[lightgray] fill between[of=K and L]; 			
	\nextgroupplot[title = Q1, ymin = 10, ymax = 15]
	\draw[dashed] (axis cs:1975,0)--(2020,0);
	\addplot[very thick] table [x expr={\thisrow{year}-1}, y=q1]{"quantiles.txt"};
	\addplot[very thick, dashed] table [x =year, y=q1]{"quantiles_MORG.txt"};
	\input{business_cycles.tex}
	\nextgroupplot[title = Q2, ymin = 12, ymax = 20]
	\draw[dashed] (axis cs:1975,0)--(2020,0);	
	\addplot[very thick] table [x expr={\thisrow{year}-1}, y=q2]{"quantiles.txt"};
	\addplot[very thick, dashed] table [x =year, y=q2]{"quantiles_MORG.txt"};
	\input{business_cycles.tex}	
	\nextgroupplot[title = Q3, ymin = 15, ymax = 32]
	\draw[dashed] (axis cs:1975,0)--(2020,0);	
	\addplot[very thick] table [x expr={\thisrow{year}-1}, y=q3]{"quantiles.txt"};
	\addplot[very thick, dashed] table [x =year, y=q3]{"quantiles_MORG.txt"};
	\input{business_cycles.tex}		
	\nextgroupplot[title = D9, ymin = 25, ymax = 45]
	\draw[dashed] (axis cs:1975,0)--(2020,0);	
	\addplot[very thick] table [x expr={\thisrow{year}-1}, y=d9]{"quantiles.txt"};
	\addplot[very thick, dashed] table [x =year, y=d9]{"quantiles_MORG.txt"};
	\input{business_cycles.tex}			
	\end{groupplot}
	\node at (plots c2r2.east) [inner sep = 10pt,anchor = north, xshift =  2cm] {\ref{grouplegend1}};  
	\end{tikzpicture}
\caption{Real hourly wage at different quantiles (measured in 2019 dollars).}
\label{WAGES1}
\end{figure}
Figure~\ref{WAGES1} presents female real hourly wage rates at selected
quantiles for both the ASEC and MORG. The shaded areas in Figure \ref{WAGES1}
indicate recessionary periods as dated by the National Bureau of Economic
Research. Quartiles are denoted by Q and deciles by D. \ Figure~\ref{WAGES1}
confirms two observations made by Lemieux (2006). First, median wages for
the ASEC and MORG are very similar. The median wage in the ASEC, despite
occasional dips, increases by 32.5 percentage points over the period
1975-2020 while the MORG shows very similar patterns with growth of 25.9\%
for 1979-2019. Second, the ASEC features more wage dispersion with
relatively lower wages at quantiles below Q2, and higher wages at quantiles
above Q2. The profiles at D1 and Q1 are similar to that at Q2 with increases
of 18.8\% and 24.1\% respectively. At Q3 and D9 there has been strong and
steady growth since 1980 with an increasing gap between each and the median
wage. Q3 increases by 40.2\% and D9 by 48.4\%. The MORG shows very similar
patterns with growth of 25.9\% for the median wage for 1979-2019, and
increases of 22.5\%, 37.1\% and 45.8\% at D1, Q3 and D9. The corresponding
increase in the ASEC median wage for the same time period is 26.8\%. The
only remarkable difference across the figures are the wages at D1 which are
clearly higher for the MORG than for the ASEC. This, as noted by Lemieux,
may reflect measurement error. However, measurement error does not explain
why the difference is greater at D1 than D9. Another cause, as we explain
below, may be the relatively lower employment rate in the MORG. This implies
that the MORG D1 is higher in the population distribution than the ASEC D1.
The difference between the data sets decreases for the MORG in 1979-1981
with a corresponding smaller decrease for the ASEC. The ASEC and the MORG
then show similar growth with the ASEC wage consistently below the MORG. As
noted by Lemieux (2006), the period 1979 to 1984 displays a sharp increase
in the residual variance in the MORG not found for the ASEC.

Figure~\ref{fig:d9d1_raw} confirms the widening wage gap with an increase in
the interquartile ratio for our sample period. For the ASEC there is an
increase of 49.4\%. For the shorter period the MORG data features an
increase of 45.0\% with the corresponding increase in the ASEC for that
period of 44.3\%. Both data sets confirm the drastic increases in female
wage inequality. The pattern associated with the interdecile ratio is
somewhat more complicated. Overall, the MORG has the larger increase but the
difference reflects wage movements for the period 1979 to 1984. The increase
in the interdecile ratio post 1984 is relatively lower for the MORG. This is
consistent with Lemieux (2006) who notes that the ASEC not only has higher
wage dispersion but also increases faster over time.

\pgfplotstableread{q3q1_females_2021.txt}{\resultsa} %
\pgfplotstableread{q3q1_females_2021_morg.txt}{\resultsb}

\pgfkeys{/pgf/number format/.cd,1000 sep = {}}

\renewcommand{\MaxBusinessCycle}{5} \renewcommand{\MinBusinessCycle}{-5}

\begin{figure}[tbp]
\centering
\begin{tikzpicture}
	\begin{groupplot}
	[
	group style = {%
		group size = 2 by 1,
		group name = plots
		, horizontal sep = 0.9cm, 
		xlabels at = edge bottom,
		ylabels at = edge left,
		x descriptions at = edge bottom
	},
	set layers,cell picture = true,
	width = 0.45\textwidth,
	height = 0.45\textwidth,
	legend columns = 1,
	xlabel = Year,
	ylabel = Ratio,
	xmin = 1975, 
	xmax = 2020,
	ymin = 1.7,
	ymax = 2.9,
	cycle list name = black white
	]
	\nextgroupplot[title = Interquartile, legend to name = grouplegend2]
	\addplot[black, very thick] table [x expr={\thisrow{year}-1}, y = stat]{\resultsa};
		\addlegendentry{ASEC}
		\addplot[black, very thick, dashed] table [x = year, y = stat]{\resultsb};
			\addlegendentry{MORG}
		\input{business_cycles.tex}
	\nextgroupplot[title = Interdecile, ymin = 2.7, ymax = 4.9]
\addplot[black, very thick] table [x expr={\thisrow{year}-1}, y = d9d1]{d9d1.txt};
\addplot[black, very thick, dashed] table [x = year, y = d9d1]{d9d1_MORG.txt};
\input{business_cycles.tex}
	\end{groupplot}
	\node at (plots c2r1.east) [inner sep=10pt,anchor=north, xshift= 1.5cm] {\ref{grouplegend2}};  
	\end{tikzpicture}
\caption{Measures of inequality.}
\label{fig:d9d1_raw}
\end{figure}

Figure \ref{LFSAVGH} presents the female employment rates and average
working weeks and hours of those employed. As explained above, the higher
employment rate for ASEC is expected. The employment rates increased
drastically during the last decades of the previous century before
decreasing slightly in the 2000s and then slightly rebounding in the most
recent years. The decreased employment since 2000 resulted in the US
employment rate at the end of the Great Recession returning to the 1980
level (see Beaudry et al. 2016). The average number of working hours per
week increased over the sample period from 35.3 to 38.3 for the ASEC and
from 35.3 to 37.8 for the MORG. Hours are highly cyclical. There is not a
large difference between the two samples although the increase in working
hours has been slightly smaller in the 1990s for the MORG. The number of
weeks also increased steeply from 42.0 to 46.2. The drop in 2020 during the
COVID19 pandemic is noteworthy.

\pgfplotstableread{lfsavgh.txt}{\lfsavgh} 
\pgfkeys{/pgf/number
	format/.cd,1000 sep = {}}

\begin{figure}[tbp]
\begin{tikzpicture}
	\begin{groupplot}[
	group style = {
		rows = 1,
		group name = plots,
		columns = 2,
		horizontal sep = 50pt,
		vertical sep = 50pt,
	},
	set layers,cell picture = true,
	legend columns = 2,
	xlabel = Year,
	width = 0.45 \textwidth,
	height = 0.45 \textwidth, 
	xmin = 1975,
	xmax = 2020,	
	ymin = 50, 
	ymax = 100,
	xtick = {1980,1990,2000,2010, 2020},	
	]
	\nextgroupplot[title = Employment rate, legend to name = grouplegend2, ylabel=Percentage]
	\addplot[black, very thick, dotted] table [x expr={\thisrow{year}-1}, y = lfs]{lfsavgh_f_ASEC.txt};
		\addlegendentry{ASEC}
	\addplot[black, very thick] table [x = year, y = lfs]{lfsavgh_f_morg.txt};
		\addlegendentry{MORG}
	\addplot[domain=1981.0:1981.6, name path = C]{100};
	\addplot[domain=1981.0:1981.6, name path = D]{-1};      
	\addplot[domain=1982.6:1983.92, name path = E]{100};
	\addplot[domain=1982.6:1983.92, name path = F]{-1};      		
	\addplot[domain=1991.6:1992.25, name path = G]{100};
	\addplot[domain=1991.6:1992.25, name path = H]{-1};   	 
	\addplot[domain=2002.25:2002.92, name path = I]{100};
	\addplot[domain=2002.25:2002.92, name path = J]{-1};   			
	\addplot[domain=2008.09:2010.5, name path = K]{100};
	\addplot[domain=2008.09:2010.5, name path = L]{-1};   		
	\addplot[lightgray] fill between[of=C and D];
	\addplot[lightgray] fill between[of=E and F];
	\addplot[lightgray] fill between[of=G and H];           
	\addplot[lightgray] fill between[of=I and J];
	\addplot[lightgray] fill between[of=K and L]; 			
	\nextgroupplot[title = Average hours/weeks, ymin = 32, ymax=50, legend to name = grouplegend3, ylabel =Hours/weeks]
	\addplot[black, very thick] table [x expr={\thisrow{year}-1}, y = weeks]{weeks_ASEC.txt};
			\addlegendentry{Weeks, ASEC}
	\addplot[black, very thick, dotted] table [x expr={\thisrow{year}-1}, y = hours]{hours_ASEC.txt};
			\addlegendentry{Hours, ASEC}
	\addplot[black, very thick, dashed] table [x = year, y = hours]{hours_MORG.txt};
			\addlegendentry{Hours, MORG}
	\addplot[domain=1981.0:1981.6, name path = C]{50};
	\addplot[domain=1981.0:1981.6, name path = D]{-1};      
	\addplot[domain=1982.6:1983.92, name path = E]{50};
	\addplot[domain=1982.6:1983.92, name path = F]{-1};      		
	\addplot[domain=1991.6:1992.25, name path = G]{50};
	\addplot[domain=1991.6:1992.25, name path = H]{-1};   	 
	\addplot[domain=2002.25:2002.92, name path = I]{50};
	\addplot[domain=2002.25:2002.92, name path = J]{-1};   			
	\addplot[domain=2008.09:2010.5, name path = K]{50};
	\addplot[domain=2008.09:2010.5, name path = L]{-1};   		
	\addplot[lightgray] fill between[of=C and D];
	\addplot[lightgray] fill between[of=E and F];
	\addplot[lightgray] fill between[of=G and H];           
	\addplot[lightgray] fill between[of=I and J];
	\addplot[lightgray] fill between[of=K and L]; 			
	\end{groupplot}
	\node at (plots c1r1.south) [inner sep=10pt,anchor=north, yshift=-5ex] {\ref{grouplegend2}};  
	\node at (plots c2r1.south) [inner sep=10pt,anchor=north, yshift=-5ex] {\ref{grouplegend3}};  
	\end{tikzpicture}
\caption{Employment rates and hours/weeks worked of employees.}
\label{LFSAVGH}
\end{figure}

Figure~\ref{WAGES1_FTFY} reveals that the wage levels of FTFY workers,
defined as working 52 weeks per year for at least 40 hours, are somewhat
higher than for all workers. Their fraction increased from 50 to 71\%. The
difference in wages diminishes at higher quantiles and, though not reported
here, becomes negligible at D9. Contrasting full-year (FY) and full-time
(FT) wages against non-FY and non-FT wages provides a similar finding. The
differences at the bottom of the distribution in comparison to the top, may
reflect that the choice of working FT and/or FY should be incorporated in
the selection model. Moreover, as non-FTFY (or non-FT/non-FY) workers are
potentially more vulnerable to interruptions it seems inappropriate to
exclude them in an evaluation of wage inequality. The wage differences
between these two working categories might also reflect selection effects.
The failure to include those who do not work FTFY means that the selection
effects in earlier studies may reflect movement from the non-FTFY to FTFY,
rather than from non-employment to FTFY. 

\begin{figure}[tbp]
\begin{tikzpicture}
	\begin{groupplot}[
	group style = {
		group size = 3 by 1,
		group name = plots,
		columns = 3,
		horizontal sep = 20pt,
		vertical sep = 2.6cm, 
		y descriptions at = edge left,
		ylabels at = edge left
	},
	set layers,cell picture = true,
	legend columns = 6,
	xlabel = Year,
	ylabel = Log hourly wages, 	
	xtick = {1980,2000,2020},
	width = 0.35\textwidth,
	height = 0.35\textwidth, 
	xmin = 1975,
	xmax = 2020,
	ymin = 2,
	ymax = 3.5
	]
	\nextgroupplot[legend to name = grouplegend2, title = Median]
	\addplot[black, very thick, dotted] table [x expr={\thisrow{year}-1}, y = q]{non_ftfy_50.txt};
	\addlegendentry{Non-FTFY}
	\addplot[black, very thick, dashed] table [x expr={\thisrow{year}-1}, y = q]{ftfy_50.txt};
		\input{business_cycles.tex}
	\addlegendentry{FTFY}
	\nextgroupplot[title = Q1]
	\addplot[black, very thick, dotted] table [x expr={\thisrow{year}-1}, y = q]{non_ftfy_25.txt};
	\addplot[black, very thick, dashed] table [x expr={\thisrow{year}-1}, y = q]{ftfy_25.txt};
		\input{business_cycles.tex}
	\nextgroupplot[title = Q3]
	\addplot[black, very thick, dotted] table [x expr={\thisrow{year}-1}, y = q]{non_ftfy_75.txt};
	\addplot[black, very thick, dashed] table [x expr={\thisrow{year}-1}, y = q]{ftfy_75.txt};
		\input{business_cycles.tex}
	\end{groupplot}
	\node at (plots c2r1.south) [inner sep=10pt,anchor=north, yshift=-5ex] {\ref{grouplegend2}};  
	\end{tikzpicture}
\caption{Percentiles of real log hourly wages by full-time full-year (FTFY)
employment status}
\label{WAGES1_FTFY}
\end{figure}

\section{Econometric analysis}

\label{model} 


\subsection{Model with Censored Selection}

\label{main_model} We consider a version of the HSM where the censoring rule
for the selection process incorporates the information on annual hours
worked rather than the binary employment/non-employment decision. The model
has the form: 
\begin{align}
Y& = g(X,E),\quad \text{if $D=1$},  \label{eq1} \\
H& = h(Z,V) ,\quad \text{if $D=1$},  \label{eq2} \\
D& =\mathbf{1}\{ h(Z,V)>\mu (Z)\},  \label{eq3}
\end{align}%
where $Y$ is the logarithm of hourly wages, $H$ is annual hours worked, $D$
is a selection or employment indicator equal to $1$ if $H$ and $Y$ are
observed or equivalently if $H$ is not censored at $\mu (Z)$, $X$ and $Z$
are vectors of observable conditioning variables, $g $, $h$ and $\mu$ are
unknown functions, and $E$ and $V$ are respectively a vector and a scalar of
potentially dependent unobservable variables with 
cumulative distribution functions (CDFs) $F_{E}$ and $F_{V}$. We assume that 
$X$ is a, not necessarily strict, subset of $Z$, i.e. $X\subseteq Z$.

We refer to equation \eqref{eq3} as the selection rule. It corresponds to
censored selection with an unobserved censoring point, that is we observe
the censoring status, $D$, but not the censoring point, $\mu(Z)$. Equations~%
\eqref{eq2}--\eqref{eq3} can be considered a reduced-form representation for
hours worked. The model is a nonparametric and nonseparable version of the
Tobit type-3 model considered by FVV, extended to incorporate an unknown
censoring threshold which is a function of $Z$. This threshold is motivated
by fixed labor costs measured in terms of hours. Individuals only work if
the desired number of hours exceeds a minimum number given by $\mu (Z)$.
Cogan (1981) shows that fixed labor costs reduce the number of individuals
working very few hours. We allow the fixed labor costs to vary by individual
and household characteristics.

Let $\perp\!\!\!\perp$ denote stochastic independence. We assume:

\begin{assumption}[Control Function]
\label{assumption:cv} $\left(E,V \right) \perp\!\!\!\perp Z $ , $V$ is a
continuously distributed random variable with strictly increasing CDF on the
support of $V$, and $v \mapsto h(Z,v)$ is strictly increasing a.s.
\end{assumption}

Without loss of generality, we can normalize $V$ to be standard uniformly
distributed. The potential dependence between $E$ and $V$ implies $Z^{\prime
}s$ independence of $E$ in the entire population does not exclude dependence
in the selected population with $D=1$. FVV showed that $Z$ is independent of 
$E$ conditional on $V$ and $D=1$ when $\mu (Z)=\mu $. Let $H^*=h(Z,V)$.
Lemma \ref{lem:cv} extends FVV's result to our model.

\begin{lemma}[Existence of Control Function]
\label{lem:cv} Under the model in \eqref{eq1}--\eqref{eq3} and Assumption %
\ref{assumption:cv}: 
\begin{equation*}
E \perp \!\!\!\perp Z\mid V,D=1.
\end{equation*}
Moreover, $V=F_{H^{\ast}\mid Z}(H\mid Z)$ for $H^*=h(Z,V)$, and $F_{H^{\ast
}\mid Z}$ is identified by 
\begin{equation*}
F_{H^{\ast }\mid Z}(h\mid z) = 1 - \pi (z)[1 - F_{H\mid Z,D}(h\mid z,1)],
\end{equation*}
where $\pi (z)=\Pr (D=1\mid Z=z)$ is the propensity score of selection.
\end{lemma}

The proof of the first statement follows from the same argument as in Lemma
1 of FVV. Thus, conditioning on $(Z,V)$ makes $D=1$ or $h(Z,V)>\mu (Z)$
deterministic. The assumption that $Z$ is independent of $(E,V)$ then proves
the result implying that $V$ is an appropriate control function.\footnote{%
This result is closely related to that of Imbens and Newey (2009), who
consider estimation and identification of a nonseparable model with a single
continuous endogenous explanatory variable.} 
The result $V=F_{H^{\ast}\mid Z}(H\mid Z)$ follows directly from the
assumption that $h$ is strictly increasing in its second argument and the
normalization on the distribution of $V$. 
Identification of $F_{H^{\ast }\mid Z}$ follows from Buchinsky and Hahn
(1998). 
In the Appendix, we propose an estimator of $F_{H^{\ast }\mid Z}$ based on
distribution regression. This estimator is an alternative to the estimators
of Buchinsky and Hahn (1998) and Chernozhukov and Hong (2002), which are
based on quantile regression.

The decompositions presented below require a wage distribution which
incorporates the value of $V$ and a statement regarding the region in which
it is identified. To proceed, we denote the support of random variables and
vectors by calligraphic letters while lower case letters in parentheses
indicate that the support is conditional on a stochastic vector taking a
particular value; e.g. $\mathcal{Z}(x)$ is the support of $Z\mid X=x$. \ We
define the Local Average Structural Function (LASF) and Local Distribution
Structural Function (LDSF) as: 
\begin{equation}
\mu (x,v):=\mathbb{E}[g(x,E)\mid V=v],\quad G(y,x,v):=\mathbb{P}[g(x,E)\leq
y\mid V=v].  \label{LDSF}
\end{equation}%
They represent the mean and distribution of $Y$ if all individuals with
control function equal to $v$ had observable characteristics equal to $x$.
An argument similar to FVV shows that: 
\begin{equation*}
\mu (x,v)=\mathbb{E}[Y\mid V=v,X=x,D=1],\quad G(y,x,v)=\mathbb{P}[Y\leq
y\mid V=v,X=x,D=1],
\end{equation*}%
provided $(x,v)\in \mathcal{XV}^{\ast }$, where: 
\begin{equation*}
\mathcal{XV}^{\ast }=\{(x,v)\in \mathcal{X}\times \mathcal{V}\mid \exists
z\in \mathcal{Z}(x):h(z,v)>\mu (z)\}.
\end{equation*}%
This set, referred to as the identification set by FVV, is identical to the
support of $(X,V)$ among the selected population. Lemma~\ref{lem:cv} implies
that the LASF and LDSF equal the mean and distribution of the observed $Y$
conditional on $(X,V)$ and that it is identified. This follows directly from 
$(E,V)\perp \!\!\!\perp Z$ and that $(x,v)\in \mathcal{XV}^{\ast }$ implies
the ability to find a $(z,v)$ combination for which $h(z,v)>\mu (z)$. We
refer to FVV for a discussion on how the size of the identified set depends
on the availability of exclusion restrictions on $Z$ with respect to $X$.

There are different candidates for $H$ in (\ref{eq2}). As the ASEC provides
both usual hours worked per week and annual hours, calculated as the product
of weeks worked last year and the usual number of hours worked per week, we
employ several alternatives. Although the usual hours per week may be the
variable in the ASEC that is closest to the hours decision in labor supply
models (Killingsworth, 1983), it may also reflect whether the job has
pre-set hours. Therefore, we employ the annual measure which incorporates
the weeks decision. As the extensive margin may capture whether an
individual has worked a positive number of hours in the past year, we also
investigate the use of the number of worked weeks. A theoretical motivation
for this measure follows from search models in which the offered wages
depend positively on the job offer arrival rate and negatively on the
separation rate (Burdett and Mortensen, 1998). As these rates also determine
the number of weeks worked, it implies a relationship between weeks worked
and wages. \label{insert:search} 
The appropriate censoring variable in the MORG is the number of hours worked
in the reference week. Note that this variable solves some of the problems
mentioned above. 
\label{main_model_end}

\subsection{Counterfactual distributions}

\label{ss:counterfactual} We consider counterfactual CDFs constructed by
integrating the LDSF with respect to different joint distributions of the
conditioning variables and control function.\footnote{%
Counterfactual means can be constructed similarly using the LASF in place of
the LDSF; see Section \ref{ss:mr}.} 
Counterfactual CDFs enable the construction of wage decompositions and
facilitate counterfactual analyses similar to those in DiNardo et al.\
(1996), \~{N}opo (2008), Fortin et al.\ (2011), and Chernozhukov et al.\
(2013). We focus on CDFs for the selected population. To simplify notation,
the superscript $s$ denotes that we condition on $D=1$. The decompositions
are based on the following representation of the CDF of the observed $Y$:%
\footnote{%
\ We refer to FVV for details.} 
\begin{equation*}
\mathbb{P}[Y\leq y \mid D=1]=:G_{Y}^{s}(y)=\int G(y,x,v)\,dF_{Z,V}^{s}(z,v),
\end{equation*}%
where: 
\begin{equation*}
F_{Z,V}^{s}(z,v)=\frac{\mathbf{1}\{h(z,v)>\mu (z)\}\,F_{Z,V}(z,v)}{\int 
\mathbf{1}\{h(z,v)>\mu (z)\}\,dF_{Z,V}(z,v)},
\end{equation*}%
denotes the joint CDF of $(Z,V)$ in the selected population and $F_{Z,V}$
denotes the joint CDF of $Z$ and $V$ in the entire population.%

The counterfactual CDFs are constructed by combining the CDFs $G$ and $%
F_{Z,V}$ with the selection rule \eqref{eq3} for different groups, each
group corresponding to a different time period or a subpopulation defined by
certain characteristics. Specifically, let $G_{t}$ be the LDSF in group $t$, 
$F_{Z_{k},V_{k}}$ be the joint CDF of $Z$ and $V$ in group~$k$, and let $%
\mathbf{1}\{ h_r(z,v)>\mu_r (z)\}$ be the selection rule in group~$r$. The
counterfactual CDF of $Y$ when $G$ is as in group~$t$, $F_{Z,V}$ is as in
group~$k$, and the selection rule is as in group~$r$ is defined as: 
\begin{equation}
G_{Y_{\langle t,k,r\rangle }}^{s}(y)=\frac{\int G_{t}(y,x,v)\,\mathbf{1}%
\{h_{r}(z,v)>\mu _{r}(z)\}\,dF_{Z_{k},V_{k}}(z,v)}{\int \mathbf{1}%
\{h_{r}(z,v)>\mu _{r}(z)\}\,dF_{Z_{k},V_{k}}(z,v)},  \label{CDF_counter}
\end{equation}
provided that the integrals are well-defined. Since the mapping $v\mapsto
h(z,v)$ is strictly monotonic, the condition $h_{r}(z,v)> \mu _{r}(z)$ in %
\eqref{CDF_counter} is equivalent to the condition: 
\begin{equation}
v>F_{H^*_{r}\mid Z}(\mu _{r}(z)\mid z)=1-\pi _{r}(z),  \label{inset}
\end{equation}%
where $F_{H^*_{r}\mid Z}(\mu _{r}(z)\mid z)$ is the probability of working
less hours than the censoring point conditional on $Z=z$ in group~$r$ and $%
\pi _{r}(z)$ is the propensity of working in that group. Given $%
G_{Y_{\langle t,k,r\rangle }}^{s}(y)$, the corresponding counterfactual
quantile function (QF) 
is: 
\begin{equation}
q_{Y_{\langle t,k,r\rangle }}^{s}(\tau )=\inf \{y\in \mathbb{R}\colon
G_{Y_{\langle t,k,r\rangle }}^{s}(y)\geq \tau \},\qquad 0 \leq \tau \leq 1.
\label{QF_counter}
\end{equation}
Under these definitions the observed CDF and QF of $Y$ for the selected
population in group~$t$ are $G_{Y_{\langle t,t,t\rangle }}^{s}$ and $%
q_{Y_{\langle t,t,t\rangle }}^{s}$ respectively.

Nonparametric identification of \eqref{CDF_counter} and \eqref{QF_counter}
depends on whether the integrals in \eqref{CDF_counter} are well defined.
They are when two conditions are met. First, if $\mathcal{Z}_{k}\subseteq 
\mathcal{Z}_{r}$, then $\pi_{r}$ is identified over all $z$ combinations in
the integral. Second, when $(\mathcal{X}\mathcal{V}_{k}\cap \mathcal{XV}%
_{r}^{\ast })\subseteq \mathcal{XV}_{t}$, then the LDSF is identified for
all combinations of $z$ on which we integrate. Here, $\mathcal{XV}_{r}^{\ast
}$ denotes the support of $(X,V)$ for the selected population in group $r$.
The identification conditions simplify when we consider two years for $q$, $%
r $, and $t$, such as $0$ and $1$, which is relevant for the decompositions.
For example, we need $\mathcal{XV}_{1}\cap \mathcal{XV}_{0}^{\ast }\subseteq 
\mathcal{XV}_{1}^{\ast }$ and $\mathcal{Z}_{1}\subseteq \mathcal{Z}_{0}$ to
identify $G_{\langle 1,1,0\rangle }$ and $\mathcal{XV}^*_{0}\subseteq 
\mathcal{XV}^*_{1}$ to identify $G_{\langle 1,0,0\rangle }$. A sufficient
condition for $\mathcal{XV}_{1}\cap \mathcal{\ XV}_{0}^{\ast }\subseteq 
\mathcal{XV}_{1}^{\ast }$ and $\mathcal{XV}^*_{0}\subseteq \mathcal{XV}%
^*_{1} $ is that the employment rates in year $0$, conditional on $X$, are
lower than those in year $1$.


Using \eqref{QF_counter}, we decompose the difference in the observed QF of $%
Y$ for the selected population between any two groups, say group~1 and
group~0, as:\footnote{%
Note that alternative orders are possible. We investigated the impact of
changes in the order on our empirical results and these were minor. Details
are available upon request.\label{footnote:order}} 
\begin{equation}
q_{Y_{\langle 1,1,1\rangle }}^{s}-q_{Y_{\langle 0,0,0\rangle }}^{s}=\underset%
{[1]}{\underbrace{[q_{Y_{\langle 1,1,1\rangle }}^{s}-q_{Y_{\langle
1,1,0\rangle }}^{s}]}}+\underset{[2]}{\underbrace{[q_{Y_{\langle
1,1,0\rangle }}^{s}-q_{Y_{\langle 1,0,0\rangle }}^{s}]}}+\underset{[3]}{%
\underbrace{[q_{Y_{\langle 1,0,0\rangle }}^{s}-q_{Y_{\langle 0,0,0\rangle
}}^{s}]}},  \label{CDFchanges}
\end{equation}%
where [1] is a selection effect that captures changes in the selection rule
given the joint distribution of $Z$ and $V$, [2] is a composition effect
that reflects changes in the joint distribution of $Z$ and $V$, and [3] is a
structural effect that reflects changes in the conditional distribution of $%
Y $ given $Z$ and $V$. These effects are relative to the base year. We
stress that this definition of the selection effect differs from the
standard definition. This is discussed in Section \ref{ss:mr}.


\subsection{Double selection mechanism}

\label{model:double}

The model can be extended to a multiple censored selection mechanism
operating through both weeks and hours. The model has the form:%
\begin{align}
Y& =g(X,E),\text{ if }h(Z,V_{H})>\mu _{H}(Z)\text{ and }w(Z,V_{W})>\mu
_{W}(Z),  \label{eq4} \\
H& =h(Z,V_{H}),\text{ if }h(Z,V_{H})>\mu _{H}(Z),  \label{eq5} \\
W& =w(Z,V_{W}),\text{ if }w(Z,V_{W})>\mu _{W}(Z),  \label{eq6}
\end{align}%
where the unobserved thresholds $\mu _{H}(Z)$ and $\mu _{W}(Z)$ are
functions of the individual's characteristics.\footnote{%
A further complication could be introduced by assuming that either the
equation \eqref{eq5} or \eqref{eq6} is the sole source of selection.} The 
distributions of $H^{\ast }:=h(Z,V_{H})$ and $W^{\ast }:=w(Z,V_{W})$
conditional on $Z$, denoted $V_{H}$ and $V_{W}$, respectively, are each
required as control functions. The analysis of this model is similar to that
above. However, it is necessary to employ both control functions to, for
example, calculate the LDSF, \emph{i.e.} 
\begin{equation*}
G(y,x,v_{H},v_{W})=\mathbb{P}(g(x,E)\leq y\mid V_{H}=v_{H},V_{W}=v_{W}).
\end{equation*}%
The identification conditions change to accommodate that the support
condition is defined over two control functions. Using the same notation as
Section \ref{ss:counterfactual}, the support requirements for the
counterfactuals are $\mathcal{ZV}_{H}\mathcal{V}_{W,k}\subseteq \mathcal{ZV}%
_{H}\mathcal{V}_{W,r}$ and $(\mathcal{XV}_{H}\mathcal{V}_{W,k}\cap \mathcal{%
XV}_{H}\mathcal{V}_{W,r}^{\ast} )\subseteq \mathcal{XV}_{H}\mathcal{V}_{W,t}$%
. We acknowledge that there are circumstances under which this model will
collapse to the single censoring mechanism case. However, as these are
somewhat obvious we do not detail them here.

\subsection{Model with ordered selection}

\label{ss:alternative_model} The models above employ control functions which
assume that the selection variable is continuous. However, both the numbers
of weeks and hours worked feature bunching at specific values (\emph{e.g.}
40 hours and 52 weeks). The following model with ordered selection
incorporates bunching: 
\begin{equation*}
\begin{split}
Y& =g(X,E)~\mbox{if}~H\geq 1, \\
H& =%
\begin{cases}
0 & \mbox{if}~V\leq \mu _{0}(Z), \\ 
1 & \mbox{if}~\mu _{0}(Z)<V\leq \mu _{1}(Z), \\ 
\vdots &  \\ 
K & \mbox{if}~V>\mu _{K-1}(Z),%
\end{cases}%
\end{split}%
\end{equation*}%
where the variables have similar interpretations as above. The main
difference between this model and those above is that it allows a discrete
distribution of $H$ at the expense of requiring separability in the
selection process. We assume $Z\perp \!\!\!\perp (E,V)$ and $V$ follows a
standard uniform distribution. 
This model is related to the ordered choice model of Heckman and Vytlacil
(2007, p. 4980), but unlike their model $g(X,E)$ does not depend on $H$. 
It can also be interpreted as an extension of Newey (2007) to multiple
ordered outcomes.

We define the identification set as:%
\begin{equation*}
\mathcal{XP}_{K}:=\{(x,p)\in \mathcal{X}\times \lbrack 0,1)\mid \exists
(h,z)\in \mathcal{HZ}(x):\mu _{h^{\prime }}(z)=p,h^{\prime }\leq h,h>0\}.
\end{equation*}%
This set collects $(x,p)$ combinations in the selected sample (i.e. $H=h>0$)
for which there is a $(h,z)$ combination in $\mathcal{HZ}(x)$ such that $\mu
_{h^{\prime }}(Z)=p$ for the propensity score of a value of $H$ smaller or
equal than the observed value $h$. For example, if $H=3$, this restriction
is satisfied when $\mu _{h^{\prime }}(z)=p$, for some $h^{\prime } \in
\{0,1,2,3\}$. We define the LDSF as in (\ref{LDSF}). We prove the following
lemma in the Appendix.

\begin{lemma}
\label{lem:ordered} Suppose that $(x,p) \in \mathcal{XP}_K$ and $\mathcal{Z}%
(x,h) = \mathcal{Z}(x)$ for all $h>0$. Then, 
\begin{equation*}
\frac{1}{1-p} \int_{p}^{1} G(y,x,v)dv = \mathbb{P}(Y \leq y \mid X=x,H > h,
\mu_h(Z) = p),
\end{equation*}
and the probability in the RHS is identified.
\end{lemma}

Lemma \ref{lem:ordered} implies $(x,p)\in \mathcal{XP}_{K}$ is also a
sufficient condition for identification as (see also Heckman and Vytlacil,
2007): 
\begin{equation*}
G(y,x,p)=-\frac{\partial }{\partial p}\int_{p}^{1}G(y,x,v)dv=\frac{\partial 
}{\partial p}\left\{ (1-p)\mathbb{P}(Y\leq y\mid X=x,H>h,\mu
_{h}(Z)=p)\right\}.
\end{equation*}%

We need additional assumptions to obtain counterfactual distributions. In
the models with continuous censoring we hold the value of the control
function constant and change the lowest value at which the individual is
participating (see \eqref{inset}). We cannot follow the same strategy here
as $V$ is not point identified. However, from the values of $H$ and $Z$ we
know that the value of $V$ is between $\mu _{H-1}(Z)$ and $\mu _{H}(Z)$.
This implies that individuals with $H=1$ have the lowest values of $V$.
Therefore, if we increase $\mu _{0}(Z)$, while leaving $V$ unchanged, some
individuals with $H=1$ would no longer participate although we do not know
who. Hence, we integrate over the distribution of $V$ and change the range
of integration accordingly. We show in the appendix that: 
\begin{equation}
G_{Y}^{s}(y) =\frac{\sum_{h=1}^{K}\int_{\mathcal{Z}}\int_{\mu
_{h-1}(z)}^{\mu _{h}(z)}G(y,x,v)dvdF_{Z}(z)}{\int_{\mathcal{Z}}(1-\mu
_{0}(z))dF_{Z}(z)},  \label{eq:alternative_model}
\end{equation}%
where $\mu_K(z) := 1$ for any $z$. This equation is comparable to equation
(7.2) of Heckman and Vytlacil, 2007). Based on (\ref{eq:alternative_model}),
the counterfactual distribution when $G$ is as in group~$t$, $F_{Z}$ is as
in group~$k$, and the selection rule is as in group~$r$ is: 
\begin{equation}
\begin{split}
G_{Y_{\langle t,k,r\rangle }}^{s}(y)& =\frac{\int_{\mathcal{Z}^{k}}\int_{\mu
_{0}^{r}(z)}^{\mu _{1}^{k}(z)}G_{t}(y,x,v)dvdF_{Z^{k}}(z)}{\int_{\mathcal{Z}%
^{k}}(1-\mu _{0}^{r}(z))dF_{Z^{k}}(z)}+ \\
& \frac{\sum_{h=2}^{K}\int_{\mathcal{Z}^{k}}\int_{\mu _{h-1}^{k}(z)}^{\mu
_{h}^{k}(z)}G_{t}(y,x,v)dvdF_{Z^{k}}(z)}{\int_{\mathcal{Z}^{k}}(1-\mu
_{0}^{r}(z))dF_{Z^{k}}(z)}.  \label{eq:cf}
\end{split}%
\end{equation}
%
The decompositions are identical to (\ref{CDFchanges}). The identification
restrictions are related to the integrals in \eqref{eq:cf}. The integral in
the numerator of the second line of \eqref{eq:cf} can be written as 
\begin{equation*}
\int_{\mathcal{Z}^{k}} \left[\int_{\mu
_{h-1}^{k}(z)}^{1}G_{t}(y,x,v)dv-\int_{\mu _{h}^{k}(z)}^{1}G_{t}(y,x,v)dv %
\right]dF_{Z^{k}}(z).
\end{equation*}
For both of these terms to be identified for any $h$, we need that $\mathcal{%
XP}_{K}^{k}\subseteq \mathcal{XP}_{K}^{t}$. For the identification of the
integral in the numerator of the first line of \eqref{eq:cf}, a similar
argument gives $(\mathcal{XP}_{K}^{r} \cap \mathcal{XP}_{K}^{k}) \subseteq 
\mathcal{XP}_{K}^{t}$. We also need that $\mathcal{Z}^{k}\subseteq \mathcal{Z%
}^{r}$ otherwise $\mu _{0}^{r}(z)$ is not identified. The identification
restrictions imply that, for example, to identify $G_{Y_{\langle
1,1,0\rangle }}^{s}$, one needs that $\mathcal{XP}_{K}^{0}\subseteq \mathcal{%
XP}_{K}^{1}$ and $\mathcal{Z}^{1}\subseteq \mathcal{Z}^{0}$. The
interpretation of these restrictions not only depends on the employment
rates between year 0 and 1 but also on whether the propensity scores in year
1 overlap those of year 0.

Despite these requirements, there is a benefit of using an ordered rather
than dichotomous selection rule. In the latter case, the restriction for $%
G_{Y_{\langle 1,1,0\rangle }}^{s}$ would have been 
that the support of the propensity scores of employment for year 1, $\mu
_{0}^{1}(Z)$, should overlap with those of year 0, $\mu _{0}^{0}(Z)$. For
ordered selection it is only necessary that one of the propensity scores in
year 1, \emph{i.e.} $\mu _{h}^{1}(Z),$ $h=1,\ldots ,K,$ overlaps with the
propensity score of employment for year 0, \emph{i.e.} $\mu _{0}^{0}(Z)$.


\subsection{Comparison with the Heckman selection model}

\label{ss:mr} The decomposition \eqref{CDFchanges} yields different effects
to those derived by MR for a parametric version of the HSM. The MR selection
effect excludes one component that we attribute to the selection effect and
includes another that we assign to the composition effect. Two other
components {are sorting effects that} cannot be separately identified from
the structural effect in nonseparable models. A common measure of a
selection effect is the difference between the average wage for the selected
population and that for the entire population. The change in the selection
effect is how this difference varies over time. We define the change in the
selection effect as the difference between the average wage in the selected
population with that which would result under a more restrictive selection
rule.

To illustrate the difference with MR, suppose that the population model in
period $t$ is the following parametric version of the HSM: 
\begin{align}
Y_{t}& =\alpha _{t}^{\prime }X_{t}+E_{t},\quad \text{if $H_{t}>0$},
\label{MR1} \\
H_{t}& =\max \{\gamma _{t}^{\prime }Z_{t}+V_{t},0\},  \label{MR2}
\end{align}%
where the first element of $X_{t}$ is the constant term, and $E_{t}$ and $%
V_{t}$ are distributed independently of $(X_{t},Z_{t})$ as bivariate normal
with zero means, variances $\sigma _{E_{t}}^{2}$ and $\sigma _{V_{t}}^{2} $,
and correlation coefficient $\rho _{t}$.\footnote{%
We present the selection rule in this censored form in order to employ our
approach although our discussion of the HSM which follows is based on the
binary rule that only $\mathbf{1(}H_{t}>0)$ is observed$.$} The
counterfactual mean of $Y$ for the selected population when the LASF is as
in group~$t$, $F_{Z,V}$ is as in group~$k$, and the selection rule is as in
group~$r$, is:\label{insert:v_changed} 
\begin{equation}
\mu _{Y_{\langle t,k,r\rangle }}^{s}=\frac{\int \mu _{t}(x,v)\,\mathbf{1}%
\{v>\Phi (-\gamma _{r}^{\prime }z)\} dF_{Z_{k},V_{k}}(z,v)}{\int \mathbf{1}%
\{v>\Phi (-\gamma _{r}^{\prime }z)\}\,dF_{Z_{k},V_{k}}(z,v)},
\label{mean_counter}
\end{equation}%
where: 
\begin{equation*}
\mu _{t}(x,v)=\alpha _{t}^{\prime }x+\rho _{t}\sigma _{E_{t}}\Phi ^{-1}(v),
\end{equation*}%
denotes the LASF in group $t$. The observed mean of $Y_{t}$ in the selected
population, integrating over $Z_{t}$, is: 
\begin{equation*}
\mu _{Y_{\langle t,t,t\rangle }}^{s}=\alpha _{t}^{\prime }\,\mathbb{E}%
[X_{t}\mid H_{t}>0]+\rho _{t}\sigma _{E_{t}}\mathbb{E}[\lambda (\gamma
_{t}^{\prime }Z_{t}/\sigma _{V_{t}})\mid H_{t}>0],
\end{equation*}%
where $\lambda$ denotes the inverse Mills ratio. We decompose the difference 
$\mu _{Y_{\langle 1,1,1\rangle }}^{s}-\mu _{Y_{\langle 0,0,0\rangle }}^{s}$
between two time periods, $t=0$ and $t=1$, into selection, composition and
structural effects.

MR define the selection effect as: 
\begin{equation}
\rho _{1}\sigma _{E_{1}}\mathbb{E}[\lambda (\gamma _{1}^{\prime
}Z_{1}/\sigma _{V_{1}})\mid H_{1}>0]-\rho _{0}\sigma _{E_{0}}\mathbb{E}%
[\lambda (\gamma _{0}^{\prime }Z_{0}/\sigma _{V_{0}})\mid H_{0}>0].
\label{eq:selection_mr}
\end{equation}%
This comprises the following four elements: 
\begin{multline}
\rho _{1}\sigma _{E_{1}}\int \left[ \lambda (\gamma _{1}^{\prime }z/\sigma
_{V_{1}})\Phi _{1}(\gamma _{1}^{\prime }z/\sigma _{V_{1}})-\lambda (\gamma
_{0}^{\prime }z/\sigma _{V_{0}})\Phi _{1}(\gamma _{0}^{\prime }z/\sigma
_{V_{0}})\right] \,dF_{Z_{1}}(z)+ \\
+\rho _{1}\sigma _{E_{1}}\left[ \int \lambda (\gamma _{0}^{\prime }z/\sigma
_{V_{0}})\Phi _{1}(\gamma _{0}^{\prime }z/\sigma
_{V_{0}})\,dF_{Z_{1}}(z)-\int \lambda (\gamma _{0}^{\prime }z/\sigma
_{V_{0}})\Phi _{0}(\gamma _{0}^{\prime }z/\sigma _{V_{0}})\,dF_{Z_{0}}(z)%
\right] + \\
+\rho _{1}\left( \sigma _{E_{1}}-\sigma _{E_{0}}\right) \mathbb{E}\left[
\lambda \left( \gamma _{0}^{\prime }Z_{1}/\sigma _{V_{0}}\right) \mid H_{0}>0%
\right] + \\
+\sigma _{E_{0}}\left( \rho _{1}-\rho _{0}\right) \mathbb{E}\left[ \lambda
\left( \gamma _{0}^{\prime }Z_{1}/\sigma _{V_{0}}\right) \mid H_{0}>0\right],
\label{MR:selection}
\end{multline}%
where $\Phi _{k}(\gamma _{r}^{\prime }z/\sigma _{V_{r}})=\Phi (\gamma
_{r}^{\prime }z/\sigma _{V_{r}})/\int \Phi (\gamma _{r}^{\prime }z/\sigma
_{V_{r}})\,dF_{Z_{k}}(z)$ is the counterfactual probability of selection in
group~$k$ when the selection rule is as in group~$r$ and $\Phi$ denotes the
standard normal CDF. 
The first two elements in \eqref{MR:selection} capture the effect of changes
over time in the observable characteristics of the selected population. The
first results from applying the selection rule from period~$1$ to period~$0$
holding the composition of period~$1$ fixed. The second captures changes in
the distribution of characteristics from period~$1$ to period~$0$. The third
element captures the effect of changes over time in the composition of the
selected population in terms of unobservables through the variance in wages.
The fourth element is a sorting effect that captures changes over time in
the composition of the selected population in terms of unobservables through
the correlation coefficient.\footnote{%
\ MR make the strong assumption that the distribution of the covariates does
not change over time, i.e., $F_{Z_{0}}=F_{Z_{1}}$, so the second component
drops out.} In our view the first element belongs to the selection effect
while the second element belongs to the composition effect as it is driven
by changes over time in the distribution of $Z$. It is also not clear why
the change in the variance of wages should be interpreted as a selection
effect as it captures factors potentially unrelated to selection. As the
final element reflects changes how unobservables are valued it is arguably a
component of the selection effect. However, it could be assigned the
interpretation of a structural effect as it captures the market value of
unobserved characteristics. We agree with the MR interpretation that the
sign of $\rho _{t}$ captures the nature of sorting.

We now present the selection, composition and structural effects for our
decomposition. Plugging the expression for $\mu _{t}(x,v)$ into %
\eqref{mean_counter} gives, after some straightforward calculations: 
\begin{equation*}
\mu _{Y_{\langle t,k,r\rangle }}^{s}=\int \left[ \alpha _{t}^{\prime }x+\rho
_{t}\sigma _{E_{t}}\lambda (\gamma _{r}^{\prime }z/\sigma _{V_{r}})\right]
\Phi _{k}(\gamma _{r}^{\prime }z/\sigma _{V_{r}})\,dF_{Z_{k}}(z).
\end{equation*}%
Our selection effect is: 
\begin{multline}
\mu _{Y_{\langle 1,1,1\rangle }}^{s}-\mu _{Y_{\langle 1,1,0\rangle
}}^{s}=\alpha _{1}^{\prime }\int x\left[ \Phi _{1}(\gamma _{1}^{\prime
}z/\sigma _{V_{1}})-\Phi _{1}(\gamma _{0}^{\prime }z/\sigma _{V_{0}})\right]
\,dF_{Z_{1}}(z)+ \\
+\rho _{1}\sigma _{E_{1}}\int \left[ \lambda (\gamma _{1}^{\prime }z/\sigma
_{V_{1}})\Phi _{1}(\gamma _{1}^{\prime }z/\sigma _{V_{1}})-\lambda (\gamma
_{0}^{\prime }z/\sigma _{V_{0}})\Phi _{1}(\gamma _{0}^{\prime }z/\sigma
_{V_{0}})\right] \,dF_{Z_{1}}(z).  \label{our:selection}
\end{multline}%
The first element on the right-hand-side of \eqref{our:selection} is the
effect on the average wage from changes in the distribution of observable
characteristics of the selected population, holding the population
distribution constant, resulting from applying the selection equation from
period~$0$ to period~$1$. It is positive if those entering the selected
population have characteristics associated with higher wages. This element
is missing in the selection effect in \eqref{MR:selection}. The second
element is the corresponding effect for the unobservable characteristics and
corresponds to the first in \eqref{MR:selection}.

Our composition effect is: 
\begin{multline}
\mu _{Y_{\langle 1,1,0\rangle }}^{s}-\mu _{Y_{\langle 1,0,0\rangle
}}^{s}=\alpha _{1}^{\prime }\left[ \int x\,\Phi _{1}(\gamma _{0}^{\prime
}z/\sigma _{V_{0}})\,dF_{Z_{1}}(z)-\int x\,\Phi _{0}(\gamma _{0}^{\prime
}z/\sigma _{V_{0}})\,dF_{Z_{0}}(z)\right] + \\
+\rho _{1}\sigma _{E_{1}}\left[ \int \lambda (\gamma _{0}^{\prime }z/\sigma
_{V_{0}})\Phi _{1}(\gamma _{0}^{\prime }z/\sigma
_{V_{0}})\,dF_{Z_{1}}(z)-\int \lambda (\gamma _{0}^{\prime }z/\sigma
_{V_{0}})\Phi _{0}(\gamma _{0}^{\prime }z/\sigma _{V_{0}})\,dF_{Z_{0}}(z)%
\right] .  \label{our:composition}
\end{multline}%
The first element on the right-hand side of \eqref{our:composition} is the
change in the average wage resulting directly from changes over time in the
distribution of the observable characteristics while the second element is
the same as the second term in (\ref{MR:selection}).

Finally, our structural effect is: 
\begin{equation}
\mu _{Y_{\langle 1,0,0\rangle }}^{s}-\mu _{Y_{\langle 0,0,0\rangle
}}^{s}=\left( \alpha _{1}-\alpha _{0}\right) ^{\prime }\mathbb{E}\left[
\left. X_{0}\right\vert H_{0}>0\right] +\left( \rho _{1}\sigma _{E_{1}}-\rho
_{0}\sigma _{E_{0}}\right) \mathbb{E}\left[ \left. \lambda \left( \gamma
_{0}^{\prime }Z_{0}/\sigma _{V_{0}}\right) \right\vert H_{0}>0\right] .
\label{our:structural}
\end{equation}%
The first element on the right-hand side of \eqref{our:structural} reflects
the impact of changes over time in the returns to observable characteristics
while the second captures the type and degree of sorting and is the same as
the {sum of the third and} fourth elements in (\ref{MR:selection}). As the
expectation involving the inverse Mills ratio is positive, its contribution
is positive when $\rho _{1}\sigma _{E_{1}}>\rho _{0}\sigma _{E_{0}}$.

Finally, consider a simple example illustrating that the two elements of the
structural effect cannot generally be identified in nonseparable models.%
\footnote{%
Chernozhukov et al. (2019) separate the structural effect from the sorting
effect using an exclusion restriction in the copula for the unobservables of
the employment and wage equation. They provide a wage decomposition with 4
components: structural, composition, selection and sorting. Compared to our
decomposition, the composition and selection effects are the same, but our
structural effect corresponds to the sum of their structural and sorting
effects.} Consider a multiplicative version of the parametric HSM obtained
by replacing \eqref{MR1} with: 
\begin{equation*}
Y_{t}=\alpha _{t}^{\prime }X_{t}\,E_{t},\quad \text{if $H_{t}>0$},
\end{equation*}%
and weaken the parametric assumption on the joint distribution of $E_{t}$
and $V_{t}$ by only requiring that $V_{t}\sim N(0,1)$ and $\mathbb{E}%
[E_{t}\mid Z_{t},H_{t}>0]=\rho _{t}\lambda (\gamma _{t}^{\prime }Z_{t})$. $%
\alpha _{t}$ and $\rho _{t}$ cannot be separately identified from the moment
condition: 
\begin{equation*}
\mathbb{E}[Y_{t}\mid Z_{t},H_{t}>0]=\alpha _{t}^{\prime }X_{t}\,\rho
_{t}\lambda (\gamma _{t}^{\prime }Z_{t}).
\end{equation*}



\section{Empirical results}

\label{results} 

\subsection{Hours equation}

We start by describing the variables included in $Z$ using the ASEC.
Following MR, we include six indicator variables for the highest educational
attainment reported. Namely, (i) 0-8 years of completed schooling, (ii) high
school dropouts, (iii) high school graduates or 12 years of schooling, (iv)
some college, (v) college, and (vi) advanced degree. We include a quartic
polynomial in potential experience and interact this with the education
levels.\footnote{%
We employ the methodology described in MR for education and potential
experience.} We use 5 indicator variables for marital status: (1) married,
(2) separated, (3) divorced, (4) widowed, and (5) never married. We include
indicator variables for Black and Hispanic and four indicator variables for
regions: northeast, midwest, south and west. Finally, we use linear terms
for the number of children aged less than 5 years interacted with the
indicator variables for marital status. For the MORG we use 5 levels of
education as the two lower categories in the ASEC are merged. The variables
Black, Hispanic, experience and region are the same. Only one indicator for
marital status is used (married or not) and we employ household size, and
its interaction with marital status, as the only household characteristic.

With the exception of the household size and composition variables, all of
the conditioning variables appear as both determinants of annual hours and
hourly wages. While one might argue that household size and composition may
affect hourly wage rates, we regard these exclusion restrictions as
reasonable and note that similar restrictions have been previously employed
(see, for example, MR). However, given their potentially contentious use we
explore the impact of not using them below. The assumption that annual hours
of work do not affect the hourly wage rate means that the variation in hours
across individuals is a source of identification.

Although our primary focus is the wage decomposition, we highlight the major
features of the hours equation estimates.\label{insert:hours_result} The
hours equation is estimated by distribution regression and we report the
marginal impact of being below a certain point in the annual hours
distribution. We do this for education level, race, marital status and for
one of the exclusion restrictions (having a child below the age of 5). We
only report the result of the latter here; see Figure \ref{fig:exclusion}.
The selected points in the hours distribution are 0, 1000 and 2000. We find
that many of the individual characteristics have an impact on the level of
annual hours worked. This is not particularly surprising given the large
literature on labor supply documenting the roles of education and marital
status on labor market participation. Perhaps what is more surprising is
that the magnitude of the impact of these variables does not appear to
change substantially over the sample period in either the ASEC or MORG data.
The exception is with respect to the exclusion restrictions which became
less important over time. This is consistent with Card and Hyslop (2021).
Note that the level of education has drastically increased over the sample
period and this has had a substantial impact on the hours distribution.

\begin{figure}[tbp]
\centering
\begin{tikzpicture}
	\begin{groupplot}[
	group style = {
		rows = 1,
		group name = plots,
		columns = 2,
		horizontal sep = 20pt,
		vertical sep = 50pt, 
		y descriptions at = edge left,
		ylabels at = edge left,
	},
	legend columns = 1,
	xlabel = Year,
	ylabel = Average derivative,	
	width = 0.45\textwidth,
	height = 0.45\textwidth, 
	xmin = 1975,
	xmax = 2020,
	ymin = 0,	
	ymax = 0.22,
	xtick = {1980, 2000,2020}
	]
	\nextgroupplot[legend to name = grouplegend2]
	\addplot[black, very thick, name path = C] table [x expr={\thisrow{year}-1}, y=h0]{iv.txt};
	\addlegendentry{0}
	\addplot[black, very thick, dashed, name path = D] table [x expr={\thisrow{year}-1}, y=h10]{iv.txt};
	\addlegendentry{1000}
	\addplot[black, very thick, dotted, name path = F] table [x expr={\thisrow{year}-1}, y=h20]{iv.txt};	
	\addlegendentry{2000}
		\addplot[domain=1981.0:1981.6, name path = C]{1};
	\addplot[domain=1981.0:1981.6, name path = D]{-1};      
	\addplot[domain=1982.6:1983.92, name path = E]{1};
	\addplot[domain=1982.6:1983.92, name path = F]{-1};      		
	\addplot[domain=1991.6:1992.25, name path = G]{1};
	\addplot[domain=1991.6:1992.25, name path = H]{-1};   	 
	\addplot[domain=2002.25:2002.92, name path = I]{1};
	\addplot[domain=2002.25:2002.92, name path = J]{-1};   			
	\addplot[domain=2008.09:2010.5, name path = K]{1};
	\addplot[domain=2008.09:2010.5, name path = L]{-1};
	\addplot[lightgray] fill between[of=C and D];
\addplot[lightgray] fill between[of=E and F];
\addplot[lightgray] fill between[of=G and H];           
\addplot[lightgray] fill between[of=I and J];
\addplot[lightgray] fill between[of=K and L]; 		   
	\end{groupplot}
	\node at (plots c1r1.east) [inner sep = 10pt,anchor = north,xshift =2cm, yshift = 5ex] {\ref{grouplegend2}};  
	\end{tikzpicture}
\caption{Average derivative of the exclusion restriction at different levels
of hours.}
\label{fig:exclusion}
\end{figure}


We also estimated models for annual weeks of work using distribution
regression and ordered models for annual weeks and annual hours using the
ASEC. The flavor of the results are very similar to those for the annual
hours equations. The individual variables which also appear in the wage
equations are statistically significant and their impact does not appear to
change a great deal over the sample period. The variables employed as
exclusion restrictions have a statistically significant impact on the level
of reported work.

\subsection{Decompositions}

\label{decompositions}

The wage equations are estimated for each year by distribution regression
over the subsample with positive hourly wages. The conditioning variables
are those in the hours equation with the exception of the household size and
composition variables. We also include the appropriate control function, its
square, and interactions between the other conditioning variables and the
control function. The decompositions require a base year. The steadily
increasing female participation rate suggests that 1975 is a reasonable
choice as it has the lowest level of participation. It is reasonable to
assume that those with a certain combination of $x$ and $v$ working in 1975
also have a positive probability of working in other years as required by
the identification conditions. We present results for the censoring
mechanism being either the number of annual hours or the number of annual
weeks for the ASEC. These are shown in Figures \ref{fig:decomp_annual} and %
\ref{fig:decomp_wks} respectively. The results for the MORG, using numbers
of hours in reference week as the censoring mechanism, are in Figure \ref%
{fig:decomp_morg}. Before presenting the decompositions we highlight that
they reflect the change in the contributions of each component over time.
For example, if the structural effects are zero this should be interpreted
as the contribution of the structural effects not changing, not that there
are no structural effects.

We start with annual hours as the censoring mechanism and Figure \ref%
{fig:decomp_annual}A presents the decomposition for the mean, which
increases by 25\% over our sample period. The total effect is driven by the
composition effect although in several instances the structural effect is
contributing. It is generally negative and small relative to the composition
effect. The contribution of the selection component is negative and small. A
negative selection component implies that females are positively selected
into employment and those who entered employment between 1975 and 2020 were
less productive than those already employed. 

\begin{figure}[tbp]
\begin{tikzpicture}
	\begin{groupplot}
	[
	group style = {
		group size = 2 by 3,
		group name = plots
		, horizontal sep = 0.5cm, 
		xlabels at = edge bottom,
		y descriptions at = edge left,
		ylabels at = edge left,
		x descriptions at = edge bottom
	},
	set layers,cell picture = true,
	width = 0.45\textwidth,
	height = 0.45\textwidth,
	legend columns = 1,
	xlabel = Year,
	ylabel = Difference,
	xmin = 1975,	
	xmax = 2020,	
	ymin = -0.15,
	ymax = 0.45,
	cycle list name = black white
	]
	\nextgroupplot[legend to name = grouplegend1, title = A. Mean]
	\draw[dashed] (axis cs:1975,0)--(2020,0);
	\addplot[very thick] table [x expr={\thisrow{year}-1}, y=selection]{"censoring_mean/src/censoring_results_50_2021.txt"};
	\addlegendentry{Selection}
	\addplot[very thick, dotted] table [x expr={\thisrow{year}-1},y=structural]{"censoring_mean/src/censoring_results_50_2021.txt"};
	\addlegendentry{Structural}
	\addplot[very thick, dashed] table[x expr={\thisrow{year}-1},y=composition]{"censoring_mean/src/censoring_results_50_2021.txt"};
	\addlegendentry{Composition}
	\addplot[very thick, double] table [x expr={\thisrow{year}-1}, y=total]{"censoring_mean/src/censoring_results_50_2021.txt"};
	\addlegendentry{Total}
	\addplot[domain=1981.0:1981.6, name path = C]{1};
	\addplot[domain=1981.0:1981.6, name path = D]{-1};      
	\addplot[domain=1982.6:1983.92, name path = E]{1};
	\addplot[domain=1982.6:1983.92, name path = F]{-1};      		
	\addplot[domain=1991.6:1992.25, name path = G]{1};
	\addplot[domain=1991.6:1992.25, name path = H]{-1};   	 
	\addplot[domain=2002.25:2002.92, name path = I]{1};
	\addplot[domain=2002.25:2002.92, name path = J]{-1};   			
	\addplot[domain=2008.09:2010.5, name path = K]{1};
	\addplot[domain=2008.09:2010.5, name path = L]{-1};   		
	\addplot[lightgray] fill between[of=C and D];
	\addplot[lightgray] fill between[of=E and F];
	\addplot[lightgray] fill between[of=G and H];           
	\addplot[lightgray] fill between[of=I and J];
	\addplot[lightgray] fill between[of=K and L]; 	
	\nextgroupplot[title = B. Median]
	\draw[dashed] (axis cs:1975,0)--(2020,0);
	\addplot[very thick] table [x expr={\thisrow{year}-1}, y=selection]{"censoring_4/src/censoring_results_50_2021.txt"};
	\addplot[very thick, dotted] table [x expr={\thisrow{year}-1}, y=structural]{"censoring_4/src/censoring_results_50_2021.txt"};
	\addplot[very thick, dashed] table [x expr={\thisrow{year}-1}, y=composition]{"censoring_4/src/censoring_results_50_2021.txt"};
	\addplot[very thick, double] table [x expr={\thisrow{year}-1}, y=total]{"censoring_4/src/censoring_results_50_2021.txt"};
	\addplot[domain=1981.0:1981.6, name path = C]{1};
	\addplot[domain=1981.0:1981.6, name path = D]{-1};      
	\addplot[domain=1982.6:1983.92, name path = E]{1};
	\addplot[domain=1982.6:1983.92, name path = F]{-1};      		
	\addplot[domain=1991.6:1992.25, name path = G]{1};
	\addplot[domain=1991.6:1992.25, name path = H]{-1};   	 
	\addplot[domain=2002.25:2002.92, name path = I]{1};
	\addplot[domain=2002.25:2002.92, name path = J]{-1};   			
	\addplot[domain=2008.09:2010.5, name path = K]{1};
	\addplot[domain=2008.09:2010.5, name path = L]{-1};   		
	\addplot[lightgray] fill between[of=C and D];
	\addplot[lightgray] fill between[of=E and F];
	\addplot[lightgray] fill between[of=G and H];           
	\addplot[lightgray] fill between[of=I and J];
	\addplot[lightgray] fill between[of=K and L]; 	
	\nextgroupplot[title = C. D1]
	\draw[dashed] (axis cs:1975,0)--(2020,0);
	\addplot[very thick] table [x expr={\thisrow{year}-1}, y=selection]{"censoring_4/src/censoring_results_10_2021.txt"};
	\addplot[very thick, dotted] table [x expr={\thisrow{year}-1}, y=structural]{"censoring_4/src/censoring_results_10_2021.txt"};
	\addplot[very thick, dashed] table [x expr={\thisrow{year}-1}, y=composition]{"censoring_4/src/censoring_results_10_2021.txt"};
	\addplot[very thick, double] table [x expr={\thisrow{year}-1}, y=total]{"censoring_4/src/censoring_results_10_2021.txt"};
	\addplot[domain=1981.0:1981.6, name path = C]{1};
	\addplot[domain=1981.0:1981.6, name path = D]{-1};      
	\addplot[domain=1982.6:1983.92, name path = E]{1};
	\addplot[domain=1982.6:1983.92, name path = F]{-1};      		
	\addplot[domain=1991.6:1992.25, name path = G]{1};
	\addplot[domain=1991.6:1992.25, name path = H]{-1};   	 
	\addplot[domain=2002.25:2002.92, name path = I]{1};
	\addplot[domain=2002.25:2002.92, name path = J]{-1};   			
	\addplot[domain=2008.09:2010.5, name path = K]{1};
	\addplot[domain=2008.09:2010.5, name path = L]{-1};   		
	\addplot[lightgray] fill between[of=C and D];
	\addplot[lightgray] fill between[of=E and F];
	\addplot[lightgray] fill between[of=G and H];           
	\addplot[lightgray] fill between[of=I and J];
	\addplot[lightgray] fill between[of=K and L]; 	
	\nextgroupplot[title = D. Q1]
	\draw[dashed] (axis cs:1975,0)--(2020,0);
	\addplot[very thick] table [x expr={\thisrow{year}-1}, y=selection]{"censoring_4/src/censoring_results_25_2021.txt"};
	\addplot[very thick, dotted] table [x expr={\thisrow{year}-1}, y=structural]{"censoring_4/src/censoring_results_25_2021.txt"};
	\addplot[very thick, dashed] table [x expr={\thisrow{year}-1}, y=composition]{"censoring_4/src/censoring_results_25_2021.txt"};
	\addplot[very thick, double] table [x expr={\thisrow{year}-1}, y=total]{"censoring_4/src/censoring_results_25_2021.txt"};
	\input{business_cycles.tex}
	\nextgroupplot[title = E. Q3]
	\draw[dashed] (axis cs:1975,0)--(2020,0);
	\addplot[very thick] table [x expr={\thisrow{year}-1}, y=selection]{"censoring_4/src/censoring_results_75_2021.txt"};
	\addplot[very thick, dotted] table [x expr={\thisrow{year}-1}, y=structural]{"censoring_4/src/censoring_results_75_2021.txt"};
	\addplot[very thick, dashed] table [x expr={\thisrow{year}-1}, y=composition]{"censoring_4/src/censoring_results_75_2021.txt"};
	\addplot[very thick, double] table [x expr={\thisrow{year}-1}, y=total]{"censoring_4/src/censoring_results_75_2021.txt"};
	\addplot[domain=1981.0:1981.6, name path = C]{1};
	\addplot[domain=1981.0:1981.6, name path = D]{-1};      
	\addplot[domain=1982.6:1983.92, name path = E]{1};
	\addplot[domain=1982.6:1983.92, name path = F]{-1};      		
	\addplot[domain=1991.6:1992.25, name path = G]{1};
	\addplot[domain=1991.6:1992.25, name path = H]{-1};   	 
	\addplot[domain=2002.25:2002.92, name path = I]{1};
	\addplot[domain=2002.25:2002.92, name path = J]{-1};   			
	\addplot[domain=2008.09:2010.5, name path = K]{1};
	\addplot[domain=2008.09:2010.5, name path = L]{-1};   		
	\addplot[lightgray] fill between[of=C and D];
	\addplot[lightgray] fill between[of=E and F];
	\addplot[lightgray] fill between[of=G and H];           
	\addplot[lightgray] fill between[of=I and J];
	\addplot[lightgray] fill between[of=K and L]; 	
	\nextgroupplot[title = F. D9]
	\draw[dashed] (axis cs:1975,0)--(2020,0);
	\addplot[very thick] table [x expr={\thisrow{year}-1}, y=selection]{"censoring_4/src/censoring_results_90_2021.txt"};
	\addplot[very thick, dotted] table [x expr={\thisrow{year}-1}, y=structural]{"censoring_4/src/censoring_results_90_2021.txt"};
	\addplot[very thick, dashed] table [x expr={\thisrow{year}-1}, y=composition]{"censoring_4/src/censoring_results_90_2021.txt"};
	\addplot[very thick, double] table [x expr={\thisrow{year}-1}, y=total]{"censoring_4/src/censoring_results_90_2021.txt"};
	\addplot[domain=1981.0:1981.6, name path = C]{1};
	\addplot[domain=1981.0:1981.6, name path = D]{-1};      
	\addplot[domain=1982.6:1983.92, name path = E]{1};
	\addplot[domain=1982.6:1983.92, name path = F]{-1};      		
	\addplot[domain=1991.6:1992.25, name path = G]{1};
	\addplot[domain=1991.6:1992.25, name path = H]{-1};   	 
	\addplot[domain=2002.25:2002.92, name path = I]{1};
	\addplot[domain=2002.25:2002.92, name path = J]{-1};   			
	\addplot[domain=2008.09:2010.5, name path = K]{1};
	\addplot[domain=2008.09:2010.5, name path = L]{-1};   		
	\addplot[lightgray] fill between[of=C and D];
	\addplot[lightgray] fill between[of=E and F];
	\addplot[lightgray] fill between[of=G and H];           
	\addplot[lightgray] fill between[of=I and J];
	\addplot[lightgray] fill between[of=K and L]; 		
	\end{groupplot}
	\node at (plots c2r1.east) [inner sep = 10pt,anchor = north, xshift =  2cm,yshift = -5ex] {\ref{grouplegend1}};  
	\end{tikzpicture}
\caption{Decompositions using the ASEC data and annual working hours.}
\label{fig:decomp_annual}
\end{figure}

\begin{figure}[tbp]
\begin{tikzpicture}
	\begin{groupplot}
	[
	group style = {
		group size = 2 by 3,
		group name = plots
		, horizontal sep = 0.5cm, 
		xlabels at = edge bottom,
		y descriptions at = edge left,
		ylabels at = edge left,
		x descriptions at = edge bottom
	},
	set layers,cell picture = true,
	width = 0.45\textwidth,
	height = 0.45\textwidth,
	legend columns = 1,
	xlabel = Year,
	ylabel = Difference,
	xmin = 1975,	
	xmax = 2020,	
	ymin = -0.15,
	ymax = 0.45,
	cycle list name = black white
	]
	\nextgroupplot[legend to name = grouplegend1, title = A. Mean]
	\draw[dashed] (axis cs:1975,0)--(2021,0);
	\addplot[very thick] table [x expr={\thisrow{year}-1}, y=selection]{"censoring_mean_wks/src/censoring_results_50_2021.txt"};
	\addlegendentry{Selection}
	\addplot[very thick, dotted] table [x expr={\thisrow{year}-1},y=structural]{"censoring_mean_wks/src/censoring_results_50_2021.txt"};
	\addlegendentry{Structural}
	\addplot[very thick, dashed] table[x expr={\thisrow{year}-1},y=composition]{"censoring_mean_wks/src/censoring_results_50_2021.txt"};
	\addlegendentry{Composition}
	\addplot[very thick, double] table [x expr={\thisrow{year}-1}, y=total]{"censoring_mean_wks/src/censoring_results_50_2021.txt"};
	\input{business_cycles.tex}
		\addlegendentry{Total}		
	\nextgroupplot[title = B. Median]
	\draw[dashed] (axis cs:1976,0)--(2021,0);
	\addplot[very thick] table [x expr={\thisrow{year}-1}, y=selection]{"empirical_quantile_us_wks_censoring/src/censoring_results_50_2021.txt"};
	\addplot[very thick, dotted] table [x expr={\thisrow{year}-1}, y=structural]{"empirical_quantile_us_wks_censoring/src/censoring_results_50_2021.txt"};
	\addplot[very thick, dashed] table [x expr={\thisrow{year}-1}, y=composition]{"empirical_quantile_us_wks_censoring/src/censoring_results_50_2021.txt"};
	\addplot[very thick, double] table [x expr={\thisrow{year}-1}, y=total]{"empirical_quantile_us_wks_censoring/src/censoring_results_50_2021.txt"};
	\input{business_cycles.tex}
	\nextgroupplot[title = C. D1]
	\draw[dashed] (axis cs:1976,0)--(2016,0);
	\addplot[very thick] table [x expr={\thisrow{year}-1}, y=selection]{"empirical_quantile_us_wks_censoring/src/censoring_results_10_2021.txt"};
	\addplot[very thick, dotted] table [x expr={\thisrow{year}-1}, y=structural]{"empirical_quantile_us_wks_censoring/src/censoring_results_10_2021.txt"};
	\addplot[very thick, dashed] table [x expr={\thisrow{year}-1}, y=composition]{"empirical_quantile_us_wks_censoring/src/censoring_results_10_2021.txt"};
	\addplot[very thick, double] table [x expr={\thisrow{year}-1}, y=total]{"empirical_quantile_us_wks_censoring/src/censoring_results_10_2021.txt"};
	\input{business_cycles.tex}
	\nextgroupplot[title = D. Q1]
	\draw[dashed] (axis cs:1976,0)--(2016,0);
	\addplot[very thick] table [x expr={\thisrow{year}-1}, y=selection]{"empirical_quantile_us_wks_censoring/src/censoring_results_25_2021.txt"};
	\addplot[very thick, dotted] table [x expr={\thisrow{year}-1}, y=structural]{"empirical_quantile_us_wks_censoring/src/censoring_results_25_2021.txt"};
	\addplot[very thick, dashed] table [x expr={\thisrow{year}-1}, y=composition]{"empirical_quantile_us_wks_censoring/src/censoring_results_25_2021.txt"};
	\addplot[very thick, double] table [x expr={\thisrow{year}-1}, y=total]{"empirical_quantile_us_wks_censoring/src/censoring_results_25_2021.txt"};
	\input{business_cycles.tex}
	\nextgroupplot[title = E. Q3]
	\draw[dashed] (axis cs:1976,0)--(2016,0);
	\addplot[very thick] table [x expr={\thisrow{year}-1}, y=selection]{"empirical_quantile_us_wks_censoring/src/censoring_results_75_2021.txt"};
	\addplot[very thick, dotted] table [x expr={\thisrow{year}-1}, y=structural]{"empirical_quantile_us_wks_censoring/src/censoring_results_75_2021.txt"};
	\addplot[very thick, dashed] table [x expr={\thisrow{year}-1}, y=composition]{"empirical_quantile_us_wks_censoring/src/censoring_results_75_2021.txt"};
	\addplot[very thick, double] table [x expr={\thisrow{year}-1}, y=total]{"empirical_quantile_us_wks_censoring/src/censoring_results_75_2021.txt"};
	\input{business_cycles.tex}
	\nextgroupplot[title = F. D9]
	\draw[dashed] (axis cs:1976,0)--(2016,0);
	\addplot[very thick] table [x expr={\thisrow{year}-1}, y=selection]{"empirical_quantile_us_wks_censoring/src/censoring_results_90_2021.txt"};
	\addplot[very thick, dotted] table [x expr={\thisrow{year}-1}, y=structural]{"empirical_quantile_us_wks_censoring/src/censoring_results_90_2021.txt"};
	\addplot[very thick, dashed] table [x expr={\thisrow{year}-1}, y=composition]{"empirical_quantile_us_wks_censoring/src/censoring_results_90_2021.txt"};
	\addplot[very thick, double] table [x expr={\thisrow{year}-1}, y=total]{"empirical_quantile_us_wks_censoring/src/censoring_results_90_2021.txt"};
	\input{business_cycles.tex}
	\end{groupplot}
	\node at (plots c2r1.east) [inner sep = 10pt,anchor = north, xshift =  2cm,yshift = -5ex] {\ref{grouplegend1}};  
	\end{tikzpicture}
\caption{Decompositions using the ASEC data and annual working weeks.}
\label{fig:decomp_wks}
\end{figure}

\begin{figure}[tbp]
\begin{tikzpicture}
	\begin{groupplot}
	[
	group style = {
		group size = 2 by 3,
		group name = plots
		, horizontal sep = 0.5cm, 
		xlabels at = edge bottom,
		y descriptions at = edge left,
		ylabels at = edge left,
		x descriptions at = edge bottom
	},
	set layers,cell picture = true,
	width = 0.45\textwidth,
	height = 0.45\textwidth,
	legend columns = 1,
	xlabel = Year,
	ylabel = Difference,
	xmin = 1979,	
	xmax = 2019,	
	ymin = -0.2,
	ymax = 0.5,
	cycle list name = black white
	]
	\nextgroupplot[legend to name = grouplegend1, title = A. Mean]
	\draw[dashed] (axis cs:1975,0)--(2020,0);
	\addplot[very thick] table [x =year, y=selection]{"censoring_results_50.txt"};
	\addlegendentry{Selection}
	\addplot[very thick, dotted] table [x=year,y=structural]{"censoring_results_50.txt"};
	\addlegendentry{Structural}
	\addplot[very thick, dashed] table[x=year,y=composition]{"censoring_results_50.txt"};
	\addlegendentry{Composition}
	\addplot[very thick, double] table [x=year, y=total]{"censoring_results_50.txt"};
	\addlegendentry{Total}
	\addplot[domain=1981.0:1981.6, name path = C]{1};
	\addplot[domain=1981.0:1981.6, name path = D]{-1};      
	\addplot[domain=1982.6:1983.92, name path = E]{1};
	\addplot[domain=1982.6:1983.92, name path = F]{-1};      		
	\addplot[domain=1991.6:1992.25, name path = G]{1};
	\addplot[domain=1991.6:1992.25, name path = H]{-1};   	 
	\addplot[domain=2002.25:2002.92, name path = I]{1};
	\addplot[domain=2002.25:2002.92, name path = J]{-1};   			
	\addplot[domain=2008.09:2010.5, name path = K]{1};
	\addplot[domain=2008.09:2010.5, name path = L]{-1};   		
	\addplot[lightgray] fill between[of=C and D];
	\addplot[lightgray] fill between[of=E and F];
	\addplot[lightgray] fill between[of=G and H];           
	\addplot[lightgray] fill between[of=I and J];
	\addplot[lightgray] fill between[of=K and L]; 	
	\nextgroupplot[title = B. Median]
	\draw[dashed] (axis cs:1976,0)--(2019,0);
	\addplot[very thick] table [x=year, y=selection]{"censoring_results_50.txt"};
	\addplot[very thick, dotted] table [x=year, y=structural]{"censoring_results_50.txt"};
	\addplot[very thick, dashed] table [x=year, y=composition]{"censoring_results_50.txt"};
	\addplot[very thick, double] table [x=year, y=total]{"censoring_results_50.txt"};
	\addplot[domain=1981.0:1981.6, name path = C]{1};
	\addplot[domain=1981.0:1981.6, name path = D]{-1};      
	\addplot[domain=1982.6:1983.92, name path = E]{1};
	\addplot[domain=1982.6:1983.92, name path = F]{-1};      		
	\addplot[domain=1991.6:1992.25, name path = G]{1};
	\addplot[domain=1991.6:1992.25, name path = H]{-1};   	 
	\addplot[domain=2002.25:2002.92, name path = I]{1};
	\addplot[domain=2002.25:2002.92, name path = J]{-1};   			
	\addplot[domain=2008.09:2010.5, name path = K]{1};
	\addplot[domain=2008.09:2010.5, name path = L]{-1};   		
	\addplot[lightgray] fill between[of=C and D];
	\addplot[lightgray] fill between[of=E and F];
	\addplot[lightgray] fill between[of=G and H];           
	\addplot[lightgray] fill between[of=I and J];
	\addplot[lightgray] fill between[of=K and L]; 	
	\nextgroupplot[title = C. D1]
	\draw[dashed] (axis cs:1976,0)--(2016,0);
	\addplot[very thick] table [x=year, y=selection]{"censoring_results_10.txt"};
	\addplot[very thick, dotted] table [x=year, y=structural]{"censoring_results_10.txt"};
	\addplot[very thick, dashed] table [x=year, y=composition]{"censoring_results_10.txt"};
	\addplot[very thick, double] table [x=year, y=total]{"censoring_results_10.txt"};
	\addplot[domain=1981.0:1981.6, name path = C]{1};
	\addplot[domain=1981.0:1981.6, name path = D]{-1};      
	\addplot[domain=1982.6:1983.92, name path = E]{1};
	\addplot[domain=1982.6:1983.92, name path = F]{-1};      		
	\addplot[domain=1991.6:1992.25, name path = G]{1};
	\addplot[domain=1991.6:1992.25, name path = H]{-1};   	 
	\addplot[domain=2002.25:2002.92, name path = I]{1};
	\addplot[domain=2002.25:2002.92, name path = J]{-1};   			
	\addplot[domain=2008.09:2010.5, name path = K]{1};
	\addplot[domain=2008.09:2010.5, name path = L]{-1};   		
	\addplot[lightgray] fill between[of=C and D];
	\addplot[lightgray] fill between[of=E and F];
	\addplot[lightgray] fill between[of=G and H];           
	\addplot[lightgray] fill between[of=I and J];
	\addplot[lightgray] fill between[of=K and L]; 	
	\nextgroupplot[title = D. Q1]
	\draw[dashed] (axis cs:1976,0)--(2016,0);
	\addplot[very thick] table [x=year, y=selection]{"censoring_results_25.txt"};
	\addplot[very thick, dotted] table [x=year, y=structural]{"censoring_results_25.txt"};
	\addplot[very thick, dashed] table [x=year, y=composition]{"censoring_results_25.txt"};
	\addplot[very thick, double] table [x=year, y=total]{"censoring_results_25.txt"};
	\addplot[domain=1981.0:1981.6, name path = C]{1};
	\addplot[domain=1981.0:1981.6, name path = D]{-1};      
	\addplot[domain=1982.6:1983.92, name path = E]{1};
	\addplot[domain=1982.6:1983.92, name path = F]{-1};      		
	\addplot[domain=1991.6:1992.25, name path = G]{1};
	\addplot[domain=1991.6:1992.25, name path = H]{-1};   	 
	\addplot[domain=2002.25:2002.92, name path = I]{1};
	\addplot[domain=2002.25:2002.92, name path = J]{-1};   			
	\addplot[domain=2008.09:2010.5, name path = K]{1};
	\addplot[domain=2008.09:2010.5, name path = L]{-1};   		
	\addplot[lightgray] fill between[of=C and D];
	\addplot[lightgray] fill between[of=E and F];
	\addplot[lightgray] fill between[of=G and H];           
	\addplot[lightgray] fill between[of=I and J];
	\addplot[lightgray] fill between[of=K and L]; 	
	\nextgroupplot[title = E. Q3]
	\draw[dashed] (axis cs:1976,0)--(2016,0);
	\addplot[very thick] table [x=year, y=selection]{"censoring_results_75.txt"};
	\addplot[very thick, dotted] table [x=year, y=structural]{"censoring_results_75.txt"};
	\addplot[very thick, dashed] table [x=year, y=composition]{"censoring_results_75.txt"};
	\addplot[very thick, double] table [x=year, y=total]{"censoring_results_75.txt"};
	\addplot[domain=1981.0:1981.6, name path = C]{1};
	\addplot[domain=1981.0:1981.6, name path = D]{-1};      
	\addplot[domain=1982.6:1983.92, name path = E]{1};
	\addplot[domain=1982.6:1983.92, name path = F]{-1};      		
	\addplot[domain=1991.6:1992.25, name path = G]{1};
	\addplot[domain=1991.6:1992.25, name path = H]{-1};   	 
	\addplot[domain=2002.25:2002.92, name path = I]{1};
	\addplot[domain=2002.25:2002.92, name path = J]{-1};   			
	\addplot[domain=2008.09:2010.5, name path = K]{1};
	\addplot[domain=2008.09:2010.5, name path = L]{-1};   		
	\addplot[lightgray] fill between[of=C and D];
	\addplot[lightgray] fill between[of=E and F];
	\addplot[lightgray] fill between[of=G and H];           
	\addplot[lightgray] fill between[of=I and J];
	\addplot[lightgray] fill between[of=K and L]; 	
	\nextgroupplot[title = F. D9]
	\draw[dashed] (axis cs:1976,0)--(2016,0);
	\addplot[very thick] table [x=year, y=selection]{"censoring_results_90.txt"};
	\addplot[very thick, dotted] table [x=year, y=structural]{"censoring_results_90.txt"};
	\addplot[very thick, dashed] table [x=year, y=composition]{"censoring_results_90.txt"};
	\addplot[very thick, double] table [x=year, y=total]{"censoring_results_90.txt"};
	\addplot[domain=1981.0:1981.6, name path = C]{1};
	\addplot[domain=1981.0:1981.6, name path = D]{-1};      
	\addplot[domain=1982.6:1983.92, name path = E]{1};
	\addplot[domain=1982.6:1983.92, name path = F]{-1};      		
	\addplot[domain=1991.6:1992.25, name path = G]{1};
	\addplot[domain=1991.6:1992.25, name path = H]{-1};   	 
	\addplot[domain=2002.25:2002.92, name path = I]{1};
	\addplot[domain=2002.25:2002.92, name path = J]{-1};   			
	\addplot[domain=2008.09:2010.5, name path = K]{1};
	\addplot[domain=2008.09:2010.5, name path = L]{-1};   		
	\addplot[lightgray] fill between[of=C and D];
	\addplot[lightgray] fill between[of=E and F];
	\addplot[lightgray] fill between[of=G and H];           
	\addplot[lightgray] fill between[of=I and J];
	\addplot[lightgray] fill between[of=K and L]; 		
	\end{groupplot}
	\node at (plots c2r1.east) [inner sep = 10pt,anchor = north, xshift =  2cm,yshift = -5ex] {\ref{grouplegend1}};  
	\end{tikzpicture}
\caption{Decompositions using the MORG data and annual working hours.}
\label{fig:decomp_morg}
\end{figure}

Figures~\ref{fig:decomp_annual}B-- \ref{fig:decomp_annual}F present the
decompositions for the 10th, 25th, 50th, 75th and 90th percentiles. A number
of important conclusions can be drawn. The large increases at each quantile
are driven by the large changes in the composition effects. This reflects
the increasing education levels of the female workforce. The large increases
from the composition effects at each quantile are somewhat offset by the
structural effects at lower quantiles and there is only clear evidence of an
economically important and positive structural effect at D9. At this
quantile almost half of the 44\% increase in real wages experienced over the
sample period is due to the structural effect. At lower quantiles the
structural effects are somewhat cyclical while the sustained increase at D9
reflects the increasing skill premia.

Figures~\ref{fig:decomp_annual}B-- \ref{fig:decomp_annual}F reveal no
indication of changes in the selection effects at Q3 and D9. Individuals
located here are likely to have had a relatively strong commitment to
employment in 1975 and there have been no substantial movements in their
hours distribution. Moreover, these individuals are less likely to incur
periods of unemployment. Lower down the wage distribution there is evidence
of small negative changes in selection during some years in our sample. At
D1 and Q1 the selection effects are negative and economically important. The
confidence intervals for these selection effects are presented in Figure~\ref%
{CI}. They indicate that for several time periods selection effects are
statistically significantly different from zero. At D1 the total real wage
increase is less than 10\% for the majority of the sample period and the
negative selection effect is frequently of the order of around 3\%.

\begin{figure}[tbp]
\begin{tikzpicture}
	\begin{groupplot}
	[
	group style = {
		group size = 2 by 3,
		group name = plots
		, horizontal sep = 0.5cm, 
		vertical sep = 2cm,
		y descriptions at = edge left,
		ylabels at = edge left,
	},
	set layers,cell picture = true,
	width = 0.45\textwidth,
	height = 0.45\textwidth,
	legend columns = 1,
	xlabel = Year,
	ylabel = Average derivative,
	xmin = 1975,	
	xmax = 2020,	
	ymin = -0.4, ymax=1.1,
	cycle list name = black white
	]
	\nextgroupplot[legend to name = grouplegend2, title = A. D1]
	\draw[dashed] (axis cs:1975,0)--(2016,0);
	\addplot[very thick] table [x expr={\thisrow{year}-1},y=v]{"fig_av/results_10.txt"};
	\addlegendentry{D1}
	\addplot[very thick, dotted] table [x expr={\thisrow{year}-1}, y=v]{"fig_av/results_25.txt"};
		\addlegendentry{Q1}
	\addplot[very thick, dashed] table [x expr={\thisrow{year}-1}, y=v]{"fig_av/results_50.txt"};
		\addlegendentry{Median}
	\addplot[very thick, dashdotted] table [x expr={\thisrow{year}-1}, y=v]{"fig_av/results_75.txt"};
		\addlegendentry{Q3}
	\addplot[thick] table [x expr={\thisrow{year}-1}, y=v]{"fig_av/results_90.txt"};
		\addlegendentry{D9}
	\addplot[thick, double] table [x expr={\thisrow{year}-1}, y=average]{"average_derivative_2021.txt"};
		\addlegendentry{Mean}
	\addplot[domain=1981.0:1981.6, name path = C]{2};
	\addplot[domain=1981.0:1981.6, name path = D]{-1};      
	\addplot[domain=1982.6:1983.92, name path = E]{2};
	\addplot[domain=1982.6:1983.92, name path = F]{-1};      		
	\addplot[domain=1991.6:1992.25, name path = G]{2};
	\addplot[domain=1991.6:1992.25, name path = H]{-1};   	 
	\addplot[domain=2002.25:2002.92, name path = I]{2};
	\addplot[domain=2002.25:2002.92, name path = J]{-1};   			
	\addplot[domain=2008.09:2010.5, name path = K]{2};
	\addplot[domain=2008.09:2010.5, name path = L]{-1};   		
	\addplot[lightgray] fill between[of=C and D];
	\addplot[lightgray] fill between[of=E and F];
	\addplot[lightgray] fill between[of=G and H];           
	\addplot[lightgray] fill between[of=I and J];
	\addplot[lightgray] fill between[of=K and L]; 	
	\end{groupplot}
		\node at (plots c1r1.east) [inner sep = 10pt,anchor = north, xshift=2cm] {\ref{grouplegend2}};  
	\end{tikzpicture}
\caption{Average derivatives at different quantiles}
\label{fig:av}
\end{figure}

We explore the form of sorting implied by these results. Given the
non-separable nature of the model, this is not straightforward although one
indication of the sorting pattern is how wages change at each quantile in
response to a change in the control function. This corresponds to the local
average derivative function of FVV evaluated at each quantile. This is
somewhat comparable to the coefficient on the selection term in the HSM. A
positive average derivative at a specific quantile suggests that the
unobservables increasing hours are positively correlated with wages at that
quantile. The results are shown in Figure \ref{fig:av}. At D1, Q1 and Q2
these estimated derivatives are positive and large for the whole sample
period. At Q3 the derivative is positive for the vast majority of the sample
period although there are instances where it is close to zero or very
slightly negative. At D9 it is negative but small in magnitude. The results
clearly support the existence of a positive relationship between $V$ and
wages at the bottom of the distribution implying positive selection which is
more important in the lower half of the wage distribution. 
This is similar to the findings of AB for female wages in the British labor
market for 1978 to 2000.

Annual hours is an economically attractive censoring mechanism as it
exploits the variation in annual hours induced both by hours and by weeks.
However, it is possible that selection operates either through hours or
weeks exclusively. We first address this issue by replacing annual hours
with annual weeks as the selection mechanism. The results from these
decompositions using the ASEC are in Figures \ref{fig:decomp_wks}A--\ref%
{fig:decomp_wks}F. Their primary feature is their similarity to those for
annual hours. This suggests that the control function from the annual hours
censoring mechanism is highly correlated with that from annual weeks despite
the differences in their respective distributions.

Now consider the decompositions for the MORG recalling that wages are
measured differently than in the ASEC and the hours measure is based on the
survey week. We implement our censored selection estimator using hours in
the survey week as the censoring mechanism noting that only a subset of the
exclusion restrictions used in the ASEC are available in the MORG. The
results are shown in Figures \ref{fig:decomp_morg}A-\ref{fig:decomp_morg}F.
As the MORG covers a different sample period, the figures look slightly
different from those for the ASEC. Nevertheless, the findings regarding the
role of selection are almost identical. The similarity across figures is
remarkable given the differences in measurement, data and exclusion
restrictions.


\pgfplotstableread{results_decomposition_females_0.1_2021.txt}{\decompa} %
\pgfplotstableread{results_decomposition_females_0.25_2021.txt}{\decompb} %
\pgfplotstableread{results_decomposition_females_0.5_2021.txt}{\decompc} %
\pgfplotstableread{results_decomposition_females_0.75_2021.txt}{\decompd}

\begin{figure}[tbp]
\begin{tikzpicture}
	\begin{groupplot}
	[
	group style = {
		group size = 2 by 2,
		group name = plots
		, horizontal sep = 0.5cm, 
		xlabels at = edge bottom,
		y descriptions at = edge left,
		ylabels at = edge left,
		x descriptions at = edge bottom
	},
	set layers,cell picture = true,
	width = 0.45\textwidth,
	height = 0.45\textwidth,
	legend columns = -1,
	xlabel = Year,
	ylabel = Selection component,
	ymin = -0.1,
	ymax = 0.1,
	ytick = {-0.2,-0.1, 0, 0.1,0.2},
	xmin = 1975,
	xmax = 2020,
	cycle list name = black white
	]
	\nextgroupplot[title = 10th percentile]
	\addplot[black, very thick] table[x expr={\thisrow{year}-1}, y = point]{\decompa};
	\addplot[black, name path = a] table[x expr={\thisrow{year}-1}, y = under]{\decompa};
	\addplot[black, name path = b] table[x expr={\thisrow{year}-1}, y = upper]{\decompa};
	\addplot[gray!30] fill between[ 
	of = a and b
	]; 
	\addplot[black, dashed, domain = 1976:2019] {0};
	\nextgroupplot[title = 25th percentile]
	\addplot[black, very thick] table[x expr={\thisrow{year}-1}, y = point]{\decompb};
	\addplot[black, name path = a] table[x expr={\thisrow{year}-1}, y = under]{\decompb};
	\addplot[black, name path = b] table[x expr={\thisrow{year}-1}, y = upper]{\decompb};
	\addplot[gray!30] fill between[ 
	of = a and b
	];
	\addplot[black, dashed, domain = 1976:2019] {0}; 
	\nextgroupplot[title = 50th percentile]
	\addplot[black, very thick] table[x expr={\thisrow{year}-1}, y = point]{\decompc};
	\addplot[black, name path = a] table[x expr={\thisrow{year}-1}, y = under]{\decompc};
	\addplot[black, name path = b] table[x expr={\thisrow{year}-1}, y = upper]{\decompc};
	\addplot[gray!30] fill between[ 
	of = a and b
	]; 
	\addplot[black, dashed, domain = 1976:2019] {0};
	\nextgroupplot[title = 75th percentile]
	\addplot[black, very thick] table[x expr={\thisrow{year}-1}, y = point]{\decompd};
	\addplot[black, name path = a] table[x expr={\thisrow{year}-1}, y = under]{\decompd};
	\addplot[black, name path = b] table[x expr={\thisrow{year}-1}, y = upper]{\decompd};
	\addplot[gray!30] fill between[ 
	of = a and b
	]; 
	\addplot[black, dashed, domain = 1976:2019] {0};
	\end{groupplot}
	\end{tikzpicture}
\caption{Selection components and associated 95\% confidence intervals for
females at various percentiles}
\label{CI}
\end{figure}

We acknowledge that although also employed by MR and MW, the use of
household composition variables as exclusion restrictions is controversial.
Accordingly, we reproduced the decompositions from the censored selection
mechanisms first excluding the household variables from both the hours and
wage equations and then including them in both equations. The model is now
only identified by the variation in the number of hours worked. We do not
find any remarkable changes from either model, in comparison to the
specification employed above, with respect to the presence or magnitude of
selection effects. The only notable difference is the presence of
occasionally larger negative selection effects at the bottom decile for the
specification which excludes the family composition variables from both
equations.

\subsection{Results of the double selection model}

Our results from Section \ref{decompositions} seem robust to the use of
either hours or weeks as the selection variable in the censored selection
model. Figures \ref{fig:decomp_both}A--\ref{fig:decomp_both}C report the
decomposition for the double selection mechanism. There continues to be no
evidence of selection above the median so we report the decompositions for
D1, Q1 and the median.

While there are some differences in these figures compared to those for
selection using only annual hours or annual weeks they are relatively small.
These results seem to suggest that the unobservables which increase
participation on any margin, such as usual weeks, usual hours, hours in
survey week, are all highly correlated. To pursue this possibility we
estimate the average derivative with respect to each of the control
functions evaluated at the mean wage as this provides some insight into the
source of selection. The derivatives, reported in Figure \ref%
{fig:average_derivative_w} in the appendix, indicate that at the mean wage
both sources of selection are important. The derivative for the weeks'
control function is the bigger of the two and the effect is relatively
constant with the exception of a notable decrease at the time of the
financial crisis. The hours' control function derivative is negative for the
earliest years of the sample before turning and remaining positive for the
remaining years. Prior to the financial crisis it increases in magnitude and
for a very short period it is the larger of the two. The high correlation
between the two control functions makes it difficult to interpret this
figure but this evidence suggests that selection operates both through the
weeks and hours decisions and the effect of each is similar.

\begin{figure}[tbp]
\begin{tikzpicture}
	\begin{groupplot}
	[
	group style = {
		group size = 2 by 2,
		group name = plots
		, horizontal sep = 0.5cm, 
		vertical sep = 2cm,
		y descriptions at = edge left,
		ylabels at = edge left,
	},
	set layers,cell picture = true,
	width = 0.4\textwidth,
	height = 0.4\textwidth,
	legend columns = 1,
	xlabel = Year,
	ylabel = Difference,
	xmin = 1975,	
	xmax = 2020,	
	ymin = -0.15,
	ymax = 0.4,
	cycle list name = black white
	]
	\nextgroupplot[title = A. D1]
	\draw[dashed] (axis cs:1975,0)--(2016,0);
	\addplot[very thick] table [x expr={\thisrow{year}-1},y=selection]{"empirical_quantile_us_both_censoring/src/censoring_results_10.txt"};
	\addplot[very thick, dotted] table [x expr={\thisrow{year}-1}, y=structural]{"empirical_quantile_us_both_censoring/src/censoring_results_10.txt"};
	\addplot[very thick, dashed] table [x expr={\thisrow{year}-1}, y=composition]{"empirical_quantile_us_both_censoring/src/censoring_results_10.txt"};
	\addplot[very thick, double] table [x expr={\thisrow{year}-1}, y=total]{"empirical_quantile_us_both_censoring/src/censoring_results_10.txt"};
	\addplot[domain=1981.0:1981.6, name path = C]{1};
	\addplot[domain=1981.0:1981.6, name path = D]{-1};      
	\addplot[domain=1982.6:1983.92, name path = E]{1};
	\addplot[domain=1982.6:1983.92, name path = F]{-1};      		
	\addplot[domain=1991.6:1992.25, name path = G]{1};
	\addplot[domain=1991.6:1992.25, name path = H]{-1};   	 
	\addplot[domain=2002.25:2002.92, name path = I]{1};
	\addplot[domain=2002.25:2002.92, name path = J]{-1};   			
	\addplot[domain=2008.09:2010.5, name path = K]{1};
	\addplot[domain=2008.09:2010.5, name path = L]{-1};   		
	\addplot[lightgray] fill between[of=C and D];
	\addplot[lightgray] fill between[of=E and F];
	\addplot[lightgray] fill between[of=G and H];           
	\addplot[lightgray] fill between[of=I and J];
	\addplot[lightgray] fill between[of=K and L]; 	
	\nextgroupplot[title = B. Q1]
	\draw[dashed] (axis cs:1975,0)--(2016,0);
	\addplot[very thick] table [x expr={\thisrow{year}-1},y=selection]{"empirical_quantile_us_both_censoring/src/censoring_results_25.txt"};
	\addplot[very thick, dotted] table [x expr={\thisrow{year}-1}, y=structural]{"empirical_quantile_us_both_censoring/src/censoring_results_25.txt"};
	\addplot[very thick, dashed] table [x expr={\thisrow{year}-1}, y=composition]{"empirical_quantile_us_both_censoring/src/censoring_results_25.txt"};
	\addplot[very thick, double] table [x expr={\thisrow{year}-1}, y=total]{"empirical_quantile_us_both_censoring/src/censoring_results_25.txt"};
	\addplot[domain=1981.0:1981.6, name path = C]{1};
	\addplot[domain=1981.0:1981.6, name path = D]{-1};      
	\addplot[domain=1982.6:1983.92, name path = E]{1};
	\addplot[domain=1982.6:1983.92, name path = F]{-1};      		
	\addplot[domain=1991.6:1992.25, name path = G]{1};
	\addplot[domain=1991.6:1992.25, name path = H]{-1};   	 
	\addplot[domain=2002.25:2002.92, name path = I]{1};
	\addplot[domain=2002.25:2002.92, name path = J]{-1};   			
	\addplot[domain=2008.09:2010.5, name path = K]{1};
	\addplot[domain=2008.09:2010.5, name path = L]{-1};   		
	\addplot[lightgray] fill between[of=C and D];
	\addplot[lightgray] fill between[of=E and F];
	\addplot[lightgray] fill between[of=G and H];           
	\addplot[lightgray] fill between[of=I and J];
	\addplot[lightgray] fill between[of=K and L]; 	
	\nextgroupplot[title = C. Median]
	\draw[dashed] (axis cs:1975,0)--(2019,0);
	\addplot[very thick] table [x expr={\thisrow{year}-1},y=selection]{"empirical_quantile_us_both_censoring/src/censoring_results_50.txt"};
	\addplot[very thick, dotted] table [x expr={\thisrow{year}-1}, y=structural]{"empirical_quantile_us_both_censoring/src/censoring_results_50.txt"};
	\addplot[very thick, dashed] table [x expr={\thisrow{year}-1}, y=composition]{"empirical_quantile_us_both_censoring/src/censoring_results_50.txt"};
	\addplot[very thick, double] table [x expr={\thisrow{year}-1}, y=total]{"empirical_quantile_us_both_censoring/src/censoring_results_50.txt"};	
	\addplot[domain=1981.0:1981.6, name path = C]{1};
	\addplot[domain=1981.0:1981.6, name path = D]{-1};      
	\addplot[domain=1982.6:1983.92, name path = E]{1};
	\addplot[domain=1982.6:1983.92, name path = F]{-1};      		
	\addplot[domain=1991.6:1992.25, name path = G]{1};
	\addplot[domain=1991.6:1992.25, name path = H]{-1};   	 
	\addplot[domain=2002.25:2002.92, name path = I]{1};
	\addplot[domain=2002.25:2002.92, name path = J]{-1};   			
	\addplot[domain=2008.09:2010.5, name path = K]{1};
	\addplot[domain=2008.09:2010.5, name path = L]{-1};   		
	\addplot[lightgray] fill between[of=C and D];
	\addplot[lightgray] fill between[of=E and F];
	\addplot[lightgray] fill between[of=G and H];           
	\addplot[lightgray] fill between[of=I and J];
	\addplot[lightgray] fill between[of=K and L]; 		
	\end{groupplot}
	\node at (plots c1r2.east) [inner sep = 10pt,anchor = north, xshift =  2.5cm] {\ref{grouplegend1}};  
	\end{tikzpicture}
\caption{Decompositions using a double selection mechanism based on ASEC
data.}
\label{fig:decomp_both}
\end{figure}

\subsection{Results of the ordered selection model}

Figure \ref{fig:decomp_bunching} reports the decomposition exercise using
the model in Section \ref{ss:alternative_model} incorporating bunching in
the selection variable. We first employ the ASEC with annual weeks as the
ordered selection rule. The separation values are 0, 10, 20, 30, 40 and 50
weeks. The decompositions are very similar to those for the continuous
selection rule. The most notable feature of the decompositions are the
selection effects at D1 and Q3. These effects are slightly larger than those
at the corresponding quantiles in the earlier figures and represent
important economic effects. There is clear evidence that the changes in
selection effects have decreased wages at this quantile in some time
periods. We also employed annual hours as the ordered selection variable
using 0, 500, 1000, 1500, 2000 and 2500 annual hours as the separation
values. These results, available from the authors upon request, are very
similar to those using ordered weeks. The selection effects are very similar
to those in Figure \ref{fig:decomp_bunching} although the allocation of the
changes across the structural and composition effects differs somewhat.

\begin{figure}[tbp]
\begin{tikzpicture}
	\begin{groupplot}
	[
	group style = {
		group size = 2 by 3,
		group name = plots
		, horizontal sep = 0.5cm, 
		xlabels at = edge bottom,
		y descriptions at = edge left,
		ylabels at = edge left,
		x descriptions at = edge bottom
	},
	set layers,cell picture = true,
	width = 0.45\textwidth,
	height = 0.45\textwidth,
	legend columns = 1,
	xlabel = Year,
	ylabel = Difference,
	xmin = 1975,	
	xmax = 2020,	
	ymin = -0.15,
	ymax = 0.5,
	cycle list name = black white
	]
	\nextgroupplot[title = A. Mean]
	\draw[dashed] (axis cs:1975,0)--(2021,0);
	\addplot[very thick] table [x=year,y=selection]{"quantile_2c_1/src/bunching_results_50.txt"};
	
	\addplot[very thick, dotted] table [x=year, y expr={\thisrow{structural}+\thisrow{hours}}]{"quantile_2c_1/src/bunching_results_50.txt"};
	\addplot[very thick, dashed] table [x=year, y=composition]{"quantile_2c_1/src/bunching_results_50.txt"};
	\addplot[very thick, double] table [x=year, y=total]{"quantile_2c_1/src/bunching_results_50.txt"};
	\input{business_cycles.tex}
	\nextgroupplot[title = B. Median]
	\draw[dashed] (axis cs:1975,0)--(2021,0);
	\addplot[very thick] table [x expr={\thisrow{year}-1},y=selection]{"quantile_2c_1/src/censoring_results_50_2021.txt"};
	\addplot[very thick, dotted] table [x expr={\thisrow{year}-1}, y=structural]{"quantile_2c_1/src/censoring_results_50_2021.txt"};
	\addplot[very thick, dashed] table [x expr={\thisrow{year}-1}, y=composition]{"quantile_2c_1/src/censoring_results_50_2021.txt"};
	\addplot[very thick, double] table [x expr={\thisrow{year}-1}, y=total]{"quantile_2c_1/src/censoring_results_50_2021.txt"};
	\input{business_cycles.tex}
	\nextgroupplot[title = C. D1]
	\draw[dashed] (axis cs:1975,0)--(2021,0);
	\addplot[very thick] table [x expr={\thisrow{year}-1},y=selection]{"quantile_2c_1/src/censoring_results_10_2021.txt"};
\addplot[very thick, dotted] table [x expr={\thisrow{year}-1}, y=structural]{"quantile_2c_1/src/censoring_results_10_2021.txt"};
\addplot[very thick, dashed] table [x expr={\thisrow{year}-1}, y=composition]{"quantile_2c_1/src/censoring_results_10_2021.txt"};
\addplot[very thick, double] table [x expr={\thisrow{year}-1}, y=total]{"quantile_2c_1/src/censoring_results_10_2021.txt"};
	\input{business_cycles.tex}
	\nextgroupplot[title = D. Q1]
	\draw[dashed] (axis cs:1975,0)--(2016,0);
	\addplot[very thick] table [x expr={\thisrow{year}-1},y=selection]{"quantile_2c_1/src/censoring_results_25_2021.txt"};
\addplot[very thick, dotted] table [x expr={\thisrow{year}-1}, y=structural]{"quantile_2c_1/src/censoring_results_25_2021.txt"};
\addplot[very thick, dashed] table [x expr={\thisrow{year}-1}, y=composition]{"quantile_2c_1/src/censoring_results_25_2021.txt"};
\addplot[very thick, double] table [x expr={\thisrow{year}-1}, y=total]{"quantile_2c_1/src/censoring_results_25_2021.txt"};
	\input{business_cycles.tex}
	\nextgroupplot[title = E. Q3]
	\draw[dashed] (axis cs:1975,0)--(2016,0);
	\addplot[very thick] table [x expr={\thisrow{year}-1},y=selection]{"quantile_2c_1/src/censoring_results_75_2021.txt"};
\addplot[very thick, dotted] table [x expr={\thisrow{year}-1}, y=structural]{"quantile_2c_1/src/censoring_results_75_2021.txt"};
\addplot[very thick, dashed] table [x expr={\thisrow{year}-1}, y=composition]{"quantile_2c_1/src/censoring_results_75_2021.txt"};
\addplot[very thick, double] table [x expr={\thisrow{year}-1}, y=total]{"quantile_2c_1/src/censoring_results_75_2021.txt"};
	\input{business_cycles.tex}
	\nextgroupplot[title = F. D9]
	\draw[dashed] (axis cs:1975,0)--(2016,0);
	\addplot[very thick] table [x expr={\thisrow{year}-1},y=selection]{"quantile_2c_1/src/censoring_results_90_2021.txt"};
\addplot[very thick, dotted] table [x expr={\thisrow{year}-1}, y=structural]{"quantile_2c_1/src/censoring_results_90_2021.txt"};
\addplot[very thick, dashed] table [x expr={\thisrow{year}-1}, y=composition]{"quantile_2c_1/src/censoring_results_90_2021.txt"};
\addplot[very thick, double] table [x expr={\thisrow{year}-1}, y=total]{"quantile_2c_1/src/censoring_results_90_2021.txt"};
	\input{business_cycles.tex}
	\end{groupplot}
	\node at (plots c2r1.east) [inner sep = 10pt,anchor = north, xshift =  2cm,yshift = -5ex] {\ref{grouplegend1}};  
	\end{tikzpicture}
\caption{Decompositions using the ASEC data and annual working weeks using
our alternative model taking account of bunching.}
\label{fig:decomp_bunching}
\end{figure}

\section{Comparison with Mulligan and Rubinstein (2008)}

\label{sec:comparison} MR examined the mean female full time wage and find
negative sorting effects in the 1970s. These become positive in the early
1980s and by the late 1990s they are economically large. The change from
negative to positive sorting is inconsistent with our findings and we
attempt to reconcile these contrasting conclusions. The appropriate figures
are in Appendix \ref{app:MR}.

We first explore the term $\rho _{t}\sigma _{E_{t}}$ in (\ref%
{eq:selection_mr}). We estimate the HSM using the MR sample and exclusion
restrictions to obtain the results in Figure \ref{fig:rho_sigma}-A. To
ensure the differences across results are not driven by their focus on FTFY
white females, we repeat the exercise including non whites and all those
reporting positive working hours. This is reported in Figure \ref%
{fig:rho_sigma}-B. These figures confirm the MR results. For both samples $%
\rho _{t}\sigma _{E_{t}}$ is negative at the beginning of the sample but
becomes positive and economically large during the sample period. The
results imply that the changes in the selection effects, as defined by MR,
are large.

Figure \ref{fig:decomp_heckman_full} presents the selection effect based on
our sample and the estimates from the HSM. The figure decomposes the total
selection effect into the four components shown in (\ref{MR:selection}). It
reveals that the movements in the selection effects are almost entirely due
to the change in $\rho _{t}$. The impact of the unobservables from the
selection equation on wages was initially negative but became positive and
subsequently large. Figure \ref{fig:decomp_heckman1_full} presents our
selection, structural and composition effect based on (\ref{our:selection})-(%
\ref{our:structural}). We find that our selection effect is small and the
majority of the wage change reflects composition effects. However, Figure %
\ref{fig:decomp_heckman1_full} is similar to Figure \ref{fig:decomp_annual}%
-A obtained via our approach. The reason for our smaller selection effect is
clear from Figure \ref{fig:decomp_heckman_full}. The figure reveals that the
second component, which from \eqref{MR:selection} is included in our
selection effect, is small and similar in size to that of Figure \ref%
{fig:decomp_annual}-A and Figure \ref{fig:decomp_heckman1_full}. This
suggests that the difference is partially due to the difference in
definitions recalling this reflects the inability to isolate the MR
selection effects in nonseparable models. We now examine whether components
of the remaining difference reflect the specific parametric form or the
source of identification employed.

That $\rho _{t}$ becomes large is not particularly controversial. However,
its change of sign has implications for the nature of sorting into the labor
market. As $\rho _{t}$ captures the mapping from unobservables in the work
equation to wages, we explore if our approach uncovers the same pattern by
evaluating once more the average derivative as presented in Figure \ref%
{fig:av}-A. While the derivative's value at the mean fluctuates over the
sample period, it does not change sign and is always consistent with
positive sorting.

This contrasts with MR. The three obvious causes are the use of the
normality assumption in the HSM, the identifying power introduced through
hours as a censoring variable in the selection equation, and the
nonseparable nature of our model. To address these issues we first estimate
the model using a parametric approach which relies on normality but which
exploits the variation in hours for identification purposes. We employ the
Vella (1993) procedure which estimates the hours equation by Tobit and
computes the generalized residual, defined for the $H>0$ observations as $%
H-Z^{\prime }\widehat{\gamma }$ where $\widehat{\gamma }$ is estimated by
Tobit, to include as the correction for selection in the wage equation. This
assumes hours do not directly effect hourly wage rates.\footnote{%
Hirsch (2005) provides empirical evidence supporting this assumption.} To
more closely correspond to the HSM we divide the generalized residual by the
estimated standard deviation of working hours, $\sigma _{V_{t}}$. The only
difference with the HSM is the use of the Tobit generalized residual rather
than the inverse Mills ratio. We plot the corresponding coefficient on the
Tobit generalized residual in Figure \ref{fig:av}-B. The coefficient on the
Tobit generalized residual also estimates $\rho _{t}\sigma _{E_{t}}.$

Two striking features are revealed in Figure \ref{fig:av}-B. First, under
normality the estimates of $\rho _{t}\sigma _{E_{t}}$ and the coefficient on
the generalized residual should be identical. However, the estimates are
very different and most importantly the coefficient estimate is always
positive. As there is no reason that departures from normality will bias the
estimates of $\rho _{t}\sigma _{E_{t}}$ and the coefficient for the Tobit
generalized residual in the same manner one could interpret the difference
in the estimates as evidence of non-normality. However, recall that the
Tobit generalized residual also exploits variation in the hours variable for
identification purposes and this could contribute to the difference in the
signs and the behavior of the two coefficients. Second, the pattern of
movement in the coefficient on the generalized residual is almost identical
to the average derivative of our control function despite the drastically
different ways in which each is computed. The two procedures are very
different but each exploits the variation in hours as a means of
identification.

While it appears that the use of the variation in hours as the source of
identification is the cause of the differences with MR, it is possible that
the departures from normality may also be responsible. The final approach we
explore is the use of the propensity score as the control function noting we
allow it to enter the wage equation in a nonseparable manner (see, for
example, Newey, 2007). The propensity score employs the exclusion
restrictions as the sole source of identification. We estimated the model
and computed the average derivative of wages with respect to the propensity
score. The results are presented in Figure \ref{fig:av}-C. This derivative
also changes sign as we move through the sample period and shows behavior
similar to $\rho _{t}$. 
We conclude that the differences in terms of the relationship between $E_{t}$
and $V_{t}$ between our results and MR are due to the use of the variation
in hours which appear to identify a different pattern of sorting.

For reconciling the difference in the results associated with the two
controls, it is necessary to examine their sources of variation. For the
subsample of workers, the variation in their values of the inverse Mills
ratio is due to the variation in $Z$. In contrast, an individual's value of
the Tobit generalized residual also exploits the value of $H.$ Consider a
case where $\rho _{t}$ is positive and there are two working individuals
with identical $Z$ but one receives a much larger positive value of $V_{t}$.
This produces a relatively larger positive value of $E_{t}$ and that
individual will have relatively higher hours and wages. In this setting the
value of the inverse Mills ratio for both individuals will be the same while
the Tobit generalized residual of the individual with the higher value of $H$
will be greater than the other. This suggests that the Tobit generalized
residual is capturing information regarding \textquotedblleft sorting" into
hours which is ignored by the inverse Mills ratio. Moreover, the inverse
mills ratio, unlike the Tobit generalized residual, is unable to explain the
variation in wages across these two individuals.

It is important to explore why $\rho _{t}$ might change sign for the models
identified solely by exclusion restrictions. A negative $\rho _{t}$ implies
that the working individuals with the lowest probabilities of participation
should have the lowest observed wages among individuals with the same
observed characteristics relevant for the wages, $X$. The
reverse is true for a positive $\rho _{t}$. We explore this by estimating a
wage regression identical to the second stage of MR while replacing the
inverse Mills ratio by a dummy variable for a child below the age of 5
years. The impact of having a \textquotedblleft young child" was negative
until 1982 at which time it turned, and remained, positive. This
corresponded to a period, also reported by Card and Hyslop (2021), in which
the magnitude of the negative impact of a \textquotedblleft young child" on
the employment decision decreased. While we acknowledge the presence of
other ongoing factors this change in the impact of \textquotedblleft young
child" could generate a change in the sign of $\rho _{t}.$ For example, in
the absence of other influences, the large positive influence of
\textquotedblleft young child" on the value of the inverse Mills ratio
combined with negative correlation between \textquotedblleft young child"
and wages would produce a negative value of $\rho _{t}.$ In contrast a
decreasing effect of \textquotedblleft young child" on participation would
produce a smaller value for the inverse Mills ratio and that, combined with
the positive correlation between \textquotedblleft young child" and wages,
would produce a positive $\rho _{t}.$ 

We highlight that we consider the above discussion as suggestive rather than
conclusive. Our objective is to consider the possible causes of the
differences in the results from the use of the two control functions. The
evidence suggests that part of the difference is due to the use of variation
in hours as a source of identification for $\rho _{t}.$ However, it is clear
that the effect of the exclusion variables on the hours and work decisions
is also an important factor in identifying $\rho _{t},$ and this has changed
over time. Related to this last issue is the validity of the exclusion
restrictions employed and how this validity has changed over the sample
period. While the use of either of the control functions should produce
consistent estimates of the sorting parameters when the model is correctly
specified, the impact under misspecification is unclear.

\section{Wage Inequality}

\label{sec:inequality}

\begin{figure}[tbp]
\centering
\begin{tikzpicture}
	\begin{groupplot}
	[
	group style = {%
		group size = 2 by 1,
		group name = plots
		, horizontal sep = 0.5cm, 
		xlabels at = edge bottom,
		y descriptions at = edge left,
		ylabels at = edge left,
		x descriptions at = edge bottom
	},
	set layers,cell picture = true,
	width = 0.45\textwidth,
	height = 0.45\textwidth,
	legend columns = 4,
	xlabel = Year,
	ylabel = Difference,
	ymin = -0.1,
	ymax = 0.4,
	xmin = 1975, 
	xmax = 2020,
	cycle list name = black white
	]
	\nextgroupplot[title = Interquartile ratio]
	\draw[dashed] (axis cs:1976,0)--(2021,0);
\addplot[very thick] table [x expr={\thisrow{year}-1}, y=selection]{"censoring_results_50_1_2021.txt"};
\addplot[very thick, dotted] table [x expr={\thisrow{year}-1}, y=structural]{"censoring_results_50_1_2021.txt"};
\addplot[very thick, dashed] table [x expr={\thisrow{year}-1}, y=composition]{"censoring_results_50_1_2021.txt"};
\addplot[very thick, double] table [x expr={\thisrow{year}-1}, y=total]{"censoring_results_50_1_2021.txt"};
			\input{business_cycles.tex}
\nextgroupplot[title = Interdecile ratio]
\draw[dashed] (axis cs:1976,0)--(2021,0);
\addplot[very thick] table [x expr={\thisrow{year}-1}, y=selection]{"censoring_results_50_2_2021.txt"};
\addplot[very thick, dotted] table [x expr={\thisrow{year}-1}, y=structural]{"censoring_results_50_2_2021.txt"};
\addplot[very thick, dashed] table [x expr={\thisrow{year}-1}, y=composition]{"censoring_results_50_2_2021.txt"};
\addplot[very thick, double] table [x expr={\thisrow{year}-1}, y=total]{"censoring_results_50_2_2021.txt"};
	\input{business_cycles.tex}
	\end{groupplot}
	\node at (plots c2r1.east) [inner sep = 10pt,anchor = north, xshift =  2cm] {\ref{grouplegend1}};  
	\end{tikzpicture}
\caption{Decompositions of the interquartile and interdecile ratio using the
ASEC.}
\label{fig:q3q1}
\end{figure}

We noted above that despite the large literatures on the impact of selection
on females' wages and the increase in female wage inequality, there are few
papers which focus on both issues. The important exceptions are AB, MW and
Blau et al. (2021), which evaluate the impact of selection by contrasting
changes in the observed levels of wage inequality with the counterfactual
levels associated with the total population working. The latter corresponds
to a participation rate of 100\%. \ We investigate the impact of selection
by evaluating changes in the distributions, and also inequality, by holding
the selection rule constant across time. While both approaches have merit we
prefer ours as participation rates do not approach 100\% in the sample
period.\footnote{Chernozhukov et al. (2019) considered both
approaches.}

\begin{figure}[tbp]
\centering
\begin{tikzpicture}
	\begin{groupplot}
	[
	group style = {%
		group size = 2 by 1,
		group name = plots
		, horizontal sep = 0.5cm, 
		xlabels at = edge bottom,
		y descriptions at = edge left,
		ylabels at = edge left,
		x descriptions at = edge bottom
	},
	set layers,cell picture = true,
	width = 0.45\textwidth,
	height = 0.45\textwidth,
	legend columns = 4,
	xlabel = Year,
	ylabel = Difference,
	ymin = -0.1,
	ymax = 0.45,
	xmin = 1979, 
	xmax = 2019,
	cycle list name = black white
	]
	\nextgroupplot[title = Interquartile ratio]
	\draw[dashed] (axis cs:1975,0)--(2020,0);
	\addplot[very thick] table [x expr={\thisrow{year}}, y=selection]{"censoring_results_q3q1.txt"};
	\addplot[very thick, dotted] table [x expr={\thisrow{year}}, y=structural]{"censoring_results_q3q1.txt""};
	\addplot[very thick, dashed] table [x expr={\thisrow{year}}, y=composition]{"censoring_results_q3q1.txt""};
	\addplot[very thick, double] table [x expr={\thisrow{year}}, y=total]{"censoring_results_q3q1.txt""};
	\input{business_cycles.tex}
	\nextgroupplot[title = Interdecile ratio]
	\draw[dashed] (axis cs:1976,0)--(2021,0);
	\addplot[very thick] table [x expr={\thisrow{year}}, y=selection]{"censoring_results_d9d1.txt""};
	\addplot[very thick, dotted] table [x expr={\thisrow{year}}, y=structural]{"censoring_results_d9d1.txt""};
	\addplot[very thick, dashed] table [x expr={\thisrow{year}}, y=composition]{"censoring_results_d9d1.txt""};
	\addplot[very thick, double] table [x expr={\thisrow{year}}, y=total]{"censoring_results_d9d1.txt""};
	\input{business_cycles.tex}
	\end{groupplot}
	\node at (plots c2r1.east) [inner sep = 10pt,anchor = north, xshift =  2cm] {\ref{grouplegend1}};  
	\end{tikzpicture}
\caption{Decompositions of the interquartile and interdecile ratio using the
MORG.}
\label{fig:q3q1_morg}
\end{figure}

We provide the decompositions of changes in inequality using the annual
hours as the censored selection variable for the ASEC data, hours in the
survey week for the MORG and annual weeks as the ordered selection variable
for the ASEC. For each of these models and selection rules we decompose the
interquartile and interdecile ratios. Those for annual hours using the ASEC
are reported in Figure \ref{fig:q3q1} and those for the MORG in Figure \ref%
{fig:q3q1_morg}. The interquartile ratio is driven by each of the
components. Neither the composition or structural effect dominates
throughout the sample period. The selection effect contributes throughout
the period and clearly increases inequality. The interdecile ratio is driven
primarily by the structural effect especially during the drastic increase at
the beginning of the sample period. The selection effect is clearly
important and frequently more important than the composition effect. For the
MORG the conclusions regarding the structural and composition effects are
similar to those for ASEC while the selection effects are slightly smaller.
This reflects the smaller selection effects at lower quantiles in the wage
decompositions (as presented in Figures \ref{fig:decomp_annual} and \ref%
{fig:decomp_morg}). The evidence for both data sets support that selection
has a modest but important impact on wage inequality that varies in
magnitude over the sample period. As the wage decompositions based on the
ordered selection rule suggested selection was more important than in the
censored selection models (see Figure \ref{fig:decomp_bunching}) we examine
now whether this carries over to the inequality decompositions. We do not
report the result but note that the evidence is similar to that for the
censored selection rule.

Our results indicate that as an increasing number of females have entered
the labor market they have reduced wages at the lower parts of the wage
distribution while having no impact on wages above Q2. This increases
measures of inequality based on ratios involving lower and upper quantiles.
Potentially, there are two reasons why selection increases inequality based
on whether either the observed or unobserved characteristics of those
participating has changed over time. However, an examination of education
levels, for example, suggests that observed characteristics have played a
minor role. In particular, we find that those with education higher than
high school degrees were more likely to participate over the whole sample
period and that this did not change over time. This suggests that our
results reflect changing unobserved characteristics. The selection effect
captures the difference between the observed wage distribution and the
counterfactual in which women participated as in 1975. Our decomposition
method presented in (\ref{CDF_counter}) imposes that this difference
captures the exit of females with lower levels of the control function.
Figure \ref{fig:average_derivative_1}, reveals a strong and positive
relationship between wages and the control function suggesting that
selection effects reflect that women entering the labor market were less
productive than observationally identical women participating in 1975.

Our evidence of positive sample selection over the whole period implies that
the decision to work is largely based on economic motivations. However, as
employment rates have increased this has seen a reduction in sorting on
economic grounds. This is consistent with the explanation provided by AB for
the U.K. labor market. This is also consistent with the results above that
the conventional household background family characteristics have become
less important in explaining participation and hours worked. The reduction
in positive sorting describes the changes in the hours distribution from the
mid 1980s to the end of the 1990s. Blau et al. (2021) argue that the booming
economy and welfare reform may have played an important role in the 1990s.
Our collective evidence suggests that post 2000 there was an increase in
positive selection. This supports the evidence in Blau et al. (2021).
Towards the end of the sample period it appears that the impact of selection
on inequality and, more generally, wages has returned to 1975 levels.

\section{Conclusions}

\label{conclusions} 

This paper documents the changes in female real wages over the period 1975
to 2020. We decompose these changes into structural, composition and
selection components by estimating a nonseparable model with selection.
Female wage growth at lower quantiles is modest although the median wage has
grown steadily. The increases at the upper quantiles for females are
substantial and reflect increasing skill premia. These changes have resulted
in a substantial increase in female wage inequality. As our sample period is
associated with large changes in the participation rates and the hours of
work of females we explore the role of changes in \textquotedblleft
selection" in wage movements. We find that the impact of these changes is to
decrease the wage growth of those at the lower quantiles with very little
evidence of selection effects at other locations in the female wage
distribution. The selection effects appear to increase wage inequality for
the period 1975 to 2000 and this reflects a reduction in the level of
positive sorting. However, post 2000 there appears to be an increase in
positive sorting and the selection effects on wages and inequality appear to
return to their 1975 levels by the end of our sample.


\subsection*{\, References}


\begin{description}
\item \textsc{Acemoglu D., and D. Autor} (2011), \textquotedblleft Skills,
tasks and technology: Implications for employment and
earnings.\textquotedblright\ In D.~Card and O.~Ashenfelter (eds.), \emph{%
Handbook of Labor Economics}, Vol.~4B: 1043--1171. Amsterdam: Elsevier
Science, North-Holland.

\item \textsc{Angrist J., Chernozhukov V., and I. Fern\`{a}ndez-Val} (2006),
\textquotedblleft Quantile regression under misspecification, with an
application to the U.S. wage structure.\textquotedblright\ \emph{Econometrica%
}, 74: 539--563.

\item \textsc{Arellano M., and S. Bonhomme S.} (2017), \textquotedblleft
Quantile selection models with an application to understanding changes in
wage inequality.\textquotedblright\ \emph{Econometrica}, 85: 1--28.

\item \textsc{Autor D.~H., Katz L.~F., and M. Kearney} (2008),
\textquotedblleft Trends in U.S.\ wage inequality: Revising the
revisionists.\textquotedblright\ \emph{Review of Economics and Statistics},
90: 300--323.

\item \textsc{Autor D.~H., Manning A., and C. Smith} (2016),
\textquotedblleft The contribution of the minimum wage to US wage inequality
over three decades: A reassessment.\textquotedblright\ \emph{American
Economic Journal: Applied Economics}, 8: 58--99.

\item \textsc{Blau, F.D., L.M. Kahn, N. Boboshko and M.L. Comey} (2021),
``The impact of selection into the labor force on the gender wage gap.'',
working paper, Cornell University.

\item \textsc{Beaudry P., Sand B.~M., and D. Green} (2016).
\textquotedblleft The great reversal in the demand for skill and cognitive
tasks.\textquotedblright\ \emph{Journal of Labor Economics}, {34}: S199--247.

\item \textsc{Bollinger C.~R., Hirsch B.~T., Hokayem C.~M., and J. Ziliak}
(2019), \textquotedblleft Trouble in the tails? What we know about earnings
nonresponse thirty years after Lillard, Smith, and Welch.\textquotedblright\ 
\emph{Journal of Political Economy}, 127: 2143--2185.

\item \textsc{Borjas G.} (1980), \textquotedblleft The relationship between
wages and weekly hours of work: The role of division
bias.\textquotedblright\ \emph{Journal of Human Resources}, 15: 409--423.

\item \textsc{Buchinsky, M. and J. Hahn} (1995), \textquotedblleft Quantile
regression model with unknown censoring point.\textquotedblright working
paper, Yale University, New Haven.

\item \textsc{Buchinsky, M. and J. Hahn} (1998), ``An alternative estimator
for the censored quantile regression model.'' \emph{Econometrica}, 66:
653--71.

\item \textsc{Burdett, K. and D.T. Mortensen} (1998), \textquotedblleft Wage
differentials, employer size and unemployment\textquotedblright , \emph{%
International Economic Review}, {39}: 257--73.


\item \textsc{Card D., and D. Hyslop} (2021), \textquotedblleft Female
earnings inequality: the changing role of family characteristics on the
extensive and intensive margins.\textquotedblright \emph{Journal of Labor
Economics}, {39}: S59--106.

\item \textsc{Chernozhukov, V., Fern\'{a}ndez-Val I., and S. Luo}{\small \ }%
(2019), "Distribution regression with sample selection, with an application
to wage decompositions in the UK"

\item \textsc{Chernozhukov, V., Fern\'{a}ndez-Val I., and B. Melly} (2013),
\textquotedblleft Inference on counterfactual
distributions.\textquotedblright\ \emph{Econometrica}, 81: 2205--68.

\item \textsc{Chernozhukov V., Fern\'{a}ndez-Val I., Newey W.K., Stouli S.,
and F. Vella} (2020), \textquotedblleft Semiparametric estimation of
structural functions in nonseparable triangular models.\textquotedblright 
\textit{Quantitative Economics}, {11}: 503--33.

\item \textsc{Chernozhukov V., and H. Hong} (2002), \textquotedblleft
Three-step censored quantile regression and extramarital affairs.
\textquotedblright , \textit{Journal of the American Statistical Association}%
, {97}, 872--82.

\item \textsc{Cogan, J.} (1981), \textquotedblleft Fixed costs and labor
supply.\textquotedblright \emph{Econometrica}, {49}:945--63.

\item \textsc{DiNardo J., Fortin N.~M., and T. Lemieux} (1996),
\textquotedblleft Labor market institutions and the distribution of wages,
1973--1992: A semiparametric approach." \emph{Econometrica}, 64: 1001--44.

\item \textsc{Fern\'{a}ndez-Val I., van Vuuren A., and F.Vella} (2021),
\textquotedblleft Nonseparable sample selection models with censored
selection rules.\textquotedblright\ \emph{Journal of Econometrics},
forthcoming.

\item \textsc{Fern\'{a}ndez-Val I., van Vuuren A., Vella F. and F.Peracchi}
(2022), "Hours worked and the U.S distribution of real annual earnings
1976--2019." \ \emph{Journal of Applied Econometrics, }forthcoming.

\item \textsc{Flood S., King M., Ruggles S., and J. Warren} (2015),
\textquotedblleft Integrated public use microdata series, Current Population
Survey: Version 4.0 [Machine-readable database].\textquotedblright\ Working
paper, University of Minnesota.


\item \textsc{Fortin N.~M., Lemieux T., and S. Firpo} (2011),
\textquotedblleft Decomposition methods in economics." In D.~Card and
O.~Ashenfelter (eds.), \emph{Handbook of Labor Economics}, Vol.~4A: 1--102.
Amsterdam: Elsevier Science, North-Holland.


\item \textsc{Heckman J.~J.} (1974), \textquotedblleft Shadow prices, market
wages and labor supply.\textquotedblright\ \emph{Econometrica}, 42: 679--94.

\item \textsc{Heckman J.~J.} (1979), \textquotedblleft Sample selection bias
as a specification error.\textquotedblright\ \emph{Econometrica}, 47:
153--61.

\item \textsc{Heckman J.~J., Stixrud J., and S. Urzua} (2006),
\textquotedblleft The effects of cognitive and noncognitive abilities on
labor market outcomes and social behavior.\textquotedblright\ \emph{Journal
of Labor Economics}, 24: 411--482.



\item \textsc{Heckman J.~J., and E. Vytlacil} (2007), \textquotedblleft
Econometric evaluation of social programs, part II\textquotedblright , in
J.J. Heckman and E.E. Leamer (eds.), \emph{Handbook of Econometrics}, 6A,
North-Holland: 4878--5143.

\item \textsc{Hirsch B. }(2005), "Why do part-time workers earn less? The
role of worker and job skills." \emph{Industrial and Labor Relations Review, 
}58: 525-551.

\item \textsc{Imbens G.~W., and W. Newey} (2009), \textquotedblleft
Identification and estimation of triangular simultaneous equations models
without additivity.\textquotedblright\ \emph{Econometrica}, 77: 1481--1512.


\item \textsc{Katz L.~F., and D. .Autor} (1999), \textquotedblleft Changes
in the wage structure and earnings inequality.\textquotedblright\ in
O.~Ashenfelter and D. Card (eds.), \emph{Handbook of Labor Economics},
Vol.~3A: 1463--1555. Amsterdam: Elsevier Science, North-Holland.

\item \textsc{Katz L.~F., and K. Murphy} (1992), \textquotedblleft Changes
in relative wages, 1963--1987: Supply and demand factors.\textquotedblright\ 
\emph{Quarterly Journal of Economics}, 107: 35--78.

\item \textsc{Killingsworth, M.R.} (1983), \textquotedblleft \emph{Labor
Supply}.\textquotedblright\ Cambridge University Press.

\item \textsc{Krueger A., and E. Posner} (2018), \textquotedblleft A
proposal for protecting low-income workers from monopsony and
corruption.\textquotedblright\ in J.~Shambaugh and R.~Nunn (eds.), \emph{%
Revitalizing Wage Growth. Policies to Get American Workers a Raise},
139--156. Washington, DC: Brookings.

\item \textsc{Lee D.~S.} (1999), \textquotedblleft Wage inequality in the
United States during the 1980s: Rising dispersion or falling minimum
wage?.\textquotedblright\ \emph{Quarterly Journal of Economics}, 114:
977--1023.

\item \textsc{Lemieux T.} (2006), \textquotedblleft Increasing residual wage
inequality: Composition effects, noisy data, or rising demand for
skill?\textquotedblright\ \emph{American Economic Review}, 96: 461--98.


\item \textsc{Maasoumi E., and L. Wang} (2019), \textquotedblleft The gender
gap between earnings distributions.\textquotedblright\ \emph{Journal of
Political Economy}, 127: 2438--2504.


\item \textsc{Meyer B.~D., Mok W.~K.~C., and J. Sullivan} (2015),
\textquotedblleft Household surveys in crisis.\textquotedblright\ \emph{%
Journal of Economic Perspectives}, 29: 199--226.

\item \textsc{Mulligan C., and Y. Rubinstein} (2008), \textquotedblleft
Selection, investment, and women's relative wages over
time.\textquotedblright\ \emph{Quarterly Journal of Economics}, 123:
1061--1110.

\item \textsc{Murphy K.~M., and R. Topel} (2016), \textquotedblleft Human
capital investment, inequality, and economic growth.\textquotedblright\ 
\emph{Journal of Labor Economics}, 34: S99--S127.

\item \textsc{Newey, W.K., }(2007), \textquotedblleft Nonparametric
continuous/discrete choice models.\textquotedblright\ \emph{International
Economic Review}, 48:1429--39.


\item \textsc{\~{N}opo, H. }(2008), \textquotedblleft Matching as a tool to
decompose wage gaps." \emph{Review of Economics and Statistics}, 90: 290--99.

\item \textsc{Olivetti, C. and B. Petrongolo} (2008), \textquotedblleft
Unequal pay or unequal employment? A cross country analysis of gender
gaps.\textquotedblright \emph{Journal of Labor Economics}, 26: 621--54.


\item \textsc{Praestgaard J., and Wellner J.} (1993), "Exchangeably weighted
bootstraps of the general empirical process", \emph{Annals of Probability},
21: 2053--2086.

\item \textsc{Vella F.} (1993), \textquotedblleft A simple estimator for
models with censored endogenous regressors.\textquotedblright\ \emph{%
International Economic Review}, 34: 441--57.

\item \textsc{Welch F.} (2000), \textquotedblleft Growth in women's relative
wages and inequality among men: One phenomenon or two?\textquotedblright\ 
\emph{American Economic Review Papers \& Proceedings}, 90: 444--49.
\end{description}



\newpage




%
%

\clearpage










\clearpage

\appendix
\linespread{1} \newpage \renewcommand{\thesubsection}{\Alph{subsection}} %
\renewcommand\thefigure{\thesubsection.\arabic{figure}} %
\setcounter{figure}{0}

\section*{Appendix}

\subsection{\protect\small \protect\bigskip Control function with unknown
censoring point} {\small We provide an estimator of $F_{H^{\ast }\mid X,Z}$
based on distribution regression. This is an alternative to Buchinsky and
Hahn (1998) and Chernozhukov and Hong (2002), which developed estimators
based on quantile regression. 
Start with a distribution regression model for $F_{H^{\ast }\mid Z}$. That
is:%
\begin{equation*}
F_{H^{\ast }\mid Z}(h\mid x)=\Lambda (r(z)^{\prime }\gamma(h)),
\end{equation*}%
where $\Lambda $ is a known link function and} ${\small r(z)}${\small \ is a 
}${\small d}_{r}${\small -dimensional vector of transformations of }${\small %
z}${\small \ with good approximating properties. We also assume a binary
response model for the propensity score of selection: 
\begin{equation*}
\pi (z)=\Lambda (p(z)^{\prime }\delta),
\end{equation*}%
where $p(z)$ is a }${\small d}_{p}${\small -dimensional vector of
transformations with good approximation properties such that: 
\begin{equation*}
\Lambda\left(r(z)^{\prime }\gamma(\mu(z))\right) \approx 1 -
\Lambda(p(z)^{\prime }\delta).
\end{equation*}%
}

{\small Let $\{Y_iD_i, H_{i},D_{i},Z_{i}\}_{i=1}^{n}$ be a random sample of $%
(HY, H,D,Z)$, where $HY$ denotes that $Y$ is only observed when $D=1$. The
proposed estimator consists of 2 steps: }

\begin{enumerate}
\item {\small Estimation of $\pi(z)$ using binary regression in the entire
sample: $\widehat{\pi}(z)=\Lambda (p(z)^{\prime }\widehat{\delta})$ where: 
\begin{equation*}
\widehat{\delta} \in \arg \max_{\delta \in \mathbb{R}^{d_{p}}}\sum_{i=1}^{n}%
\left\{ D_{i}\log \Lambda (p(Z_{i})^{\prime }\delta)+(1-D_{i})\log
[1-\Lambda (p(Z_{i})^{\prime }\delta)]\right\} .
\end{equation*}%
}

\item {\small Estimation of $F_{H^{\ast }\mid Z}$ by distribution regression
with sample selection correction in the selected sample: $\widehat
F_{H^{\ast }\mid Z}(h \mid z) = \Lambda(r(z)^{\prime }\widehat \gamma(h) )$
where: 
\begin{multline*}
\widehat{\gamma}(h)\in \arg \max_{\gamma \in \mathbb{R}^{d_{r}}}%
\sum_{i=1}^{n}D_{i}\left\{ 1(H_{i}\leq h)\log \left[ \Lambda(r(Z_{i})^{%
\prime }\gamma)+\widehat{\pi }(Z_{i})-1\right] \right. \\
\left. +1(H_{i}>h)\log \left[ 1-\Lambda(r(Z_{i})^{\prime }\gamma )\right]
\right\}.
\end{multline*}
}
\end{enumerate}


\subsection{\protect\small Estimation of wage equation}

{\small We estimate the LASF $\mu (x,v)$ and the LDSF $G(y,x,v)$ for each
group using flexibly parametrized ordinary least squares (OLS) and
distribution regressions, where the unknown control function is replaced by
its estimator $\widehat{V}_{i} = \widehat F_{H^{\ast }\mid Z}(H_i \mid Z_i) $
from the previous step. For reasons explained in FVV, we estimate over a
sample trimmed with respect to the censoring variable $H$. We employ the
following trimming indicator among the selected sample: 
\begin{equation*}
T=\mathbf{1}\{0<H\leq \overline{h}\}
\end{equation*}%
for some $\overline{h}\in (0,\infty )$ such that $\mathbb{P}(T=1)>0$. The
estimator of the LASF is $\widehat{\mu }(x,v)=w(x,v)^{\prime }\widehat{\beta 
}$, where $w(x,v)$ is a $d_{w}$-dimensional vector of transformations of $%
(x,v)$ with good approximating properties, and $\widehat{\beta }$ is the OLS
estimator: 
\begin{equation*}
\widehat{\beta }=\left[ \sum_{i=1}^{n}T_{i}\,\widehat{W}_{i}\widehat{W}%
_{i}^{\prime }\right] ^{-1}\sum_{i=1}^{n}T_{i}\,\widehat{W}_{i}Y_{i},
\end{equation*}%
where $\widehat{W}_{i}=w(X_{i},\widehat{V}_{i})$. The estimator of the LDSF
is $\widehat{G}(y,x,v)=\Lambda (w(x,v)^{\prime }\widehat{\beta }(y))$, where 
$y\in \mathbb{R}$ and $\widehat{\beta }(y)$ is the logistic distribution
regression estimator: 
\begin{equation*}
\widehat{\beta }(y)=\arg \max_{b\in \mathbb{R}^{d_{w}}}\sum_{i=1}^{n}T_{i}%
\left[ \mathbf{1}\{Y_{i}\leq y\}\log \Lambda (\widehat{W}_{i}^{\prime }b))+%
\mathbf{1}\{Y_{i}>y\}\log \Lambda (-\widehat{W}_{i}^{\prime }b))\right] .
\end{equation*}
}

{\small Finally, in the third step we use \eqref{inset} to estimate the
counterfactual CDF \eqref{CDF_counter} by: 
\begin{equation*}
\widehat{G}_{Y_{\langle t,k,r\rangle }}^{s}(y)=\frac{1}{n_{kr}^{s}}\sum_{i=1}^{n}\Lambda (\widehat{W}_{i}^{\prime }\widehat{\beta }_{t}(y))\,\mathbf{1}\{\widehat{V_{i}}> 1 - \widehat{\pi }_{r}(Z_i)\},
\end{equation*} where the average is taken over the sample values of $%
\widehat{V}_{i}$ and $Z_{i}$ in group~$k$, 
$n_{kr}^{s}=\sum_{i=1}^{n}\mathbf{1}\{\widehat{V}_{i}> 1 -
\widehat{\pi }_{r}(Z_i)\}$, $\widehat{\beta }_{t}(y)$ is the logistic
distribution regression estimator for group~$t$ from the second step, and $%
\widehat{\pi }_{r}(z)$ is the estimator of the propensity score of selection
for group~$r$ from the first step. Given $\widehat{G}_{Y_{\langle
t,k,r\rangle }}^{s}$, we estimate the counterfactual QF \eqref{QF_counter}
by: 
\begin{equation*}
\widehat{q}_{Y_{\langle t,k,r\rangle }}^{s}(\tau )= \int_{0}^{\infty} 
\boldsymbol{1}\{ \widehat{G}_{Y_{\langle t,k,r\rangle }}^{s}(y) \leq \tau \}
dy - \int_{-\infty}^0 \boldsymbol{1}\{ \widehat{G}_{Y_{\langle t,k,r\rangle
}}^{s}(y) > \tau \} dy.
\end{equation*}%
}{\small Following FVV, inference is based on the weighted bootstrap
(Praestgaard and Wellner 1993). 
This method obtains the bootstrap version of the estimator of interest by
repeating all the estimation steps including sampling weights drawn from a
nonnegative distribution with mean and variance equal to one (e.g., standard
exponential). 
}

\subsection{{\protect\small Proof of Lemma \protect\ref{lem:ordered}}}
{\small 
\begin{equation*}
\begin{split}
\frac{1}{1-p}\int_{p}^{1}G(y,x,v)dv& =\frac{1}{1-p}\int_{p}^{1}\mathbb{P}%
(g(x,E)\leq y\mid V=v)dv \\
& =\int_{p}^{1}\mathbb{P}(g(x,E)\leq y\mid V=v)dF_{V\mid V>p}(v) \\
& =\mathbb{P}(g(x,E)\leq y\mid V>\mu _{h}(z),Z=z,\mu _{h}(Z)=p) \\
& =\mathbb{P}(Y\leq y\mid H>h,X=x,\mu _{h}(Z)=p),
\end{split}%
\end{equation*}%
The first equality is definition. The second equality uses $V\sim U(0,1)$
and the third equality uses independence of $(E,V)$ and $Z$. The final
equality uses the definitions of $Y$ and $H$ and is identified because $%
(x,p)\in \mathcal{XP}_{K}$. }

\subsection{{\protect\small Derivations of equation 
\eqref{eq:alternative_model}}}

Adapting the representation of the distribution of the observed $Y$ in Section \ref{ss:counterfactual}  to the ordered selection rules yields
\begin{multline*}
G_{Y}^{s}(y)= \frac{\int_{\mathcal{Z}^{k}} \int G(y,x,v) \mathbf{1}\{v>\mu_0(z)\} dv dF_{Z}(z)}{\int_{\mathcal{Z}^{k}} \int 
\mathbf{1}\{v>\mu_0(z)\}\,dv dF_{Z}(z)} = \\
\frac{\sum_{h=1}^K \int_{\mathcal{Z}^{k}} \int G(y,x,v) \mathbf{1}\{\mu_{h-1}(z) < v \leq \mu_{h}(z)\} dv dF_{Z}(z)}{\int_{\mathcal{Z}^{k}} \int 
\mathbf{1}\{v>\mu_0(z)\}\,dv dF_{Z}(z)},
\end{multline*}where the second equality uses that the interval $(\mu_0(z), 1]$ is the union of the disjoint intervals $(\mu_{h-1}(z), \mu_{h}(z)]$, $h = 1,\ldots,K$.

\newpage

\subsection{\protect\small Figures}

\begin{figure}[h!]
\centering
\begin{tikzpicture}
	\begin{groupplot}
	[
	group style = {%
		group size = 1 by 1,
		group name = plots
		, horizontal sep = 1cm, 
		xlabels at = edge bottom,
		ylabels at = edge left,
		x descriptions at = edge bottom
	},
	set layers,cell picture = true,
	width = 0.45\textwidth,
	height = 0.45\textwidth,
	legend columns = 4,
	xlabel = Year,
	ylabel = Average derivative,
	xmin = 1975, 
	xmax = 2020,
	cycle list name = black white
	]
	\nextgroupplot[legend to name = grouplegend1, ymin = -0.2, ymax = 0.5]
	\draw[dashed] (axis cs:1975,0)--(2020,0);
	\addplot[very thick, dashed] table [x expr={\thisrow{year}-1}, y=hours]{"average_der_w.txt"};
	\addlegendentry{Hours}
	\addplot[very thick] table [x expr={\thisrow{year}-1}, y=weeks]{"average_der_w.txt"};
	\addlegendentry{Weeks}
	\input{business_cycles.tex}
	\end{groupplot}
	\node at (plots c1r1.south) [inner sep = 10pt,anchor = north, xshift =  2cm,yshift = -5ex] {\ref{grouplegend1}};  
	\end{tikzpicture}
\caption{Average derivative}
\label{fig:average_derivative_w}
\end{figure}

{\small \label{app:MR} 
}

{\small 
\begin{figure}[h!]
\begin{center}
{\small 
\begin{tikzpicture}
	\begin{groupplot}
	[
	group style={%
		group size = 2 by 1,
		group name=plots
		, horizontal sep=1.25cm, 
		xlabels at=edge bottom,
		ylabels at=edge left,
		x descriptions at=edge bottom
	},
	set layers,cell picture=true,
	width=0.45\textwidth,
	height=0.45\textwidth,
	legend columns=-1,
	xlabel = Year,
	ylabel = $\rho_t \sigma_{E_t}$,
	xmin = 1975, 
	xmax = 2020,
	ymax = 0.4,
	ymin = -0.2,
	cycle list name=black white
	]
	\nextgroupplot[legend to name=grouplegend1, title=A. Only white FTFY females]
	\addplot[black, very thick] table[x expr={\thisrow{year}-1}, y = selection]{results_MR_4a.txt};
	\addplot[domain=1981.0:1981.6, name path = C]{1};
	\addplot[domain=1981.0:1981.6, name path = D]{-1};      
	\addplot[domain=1982.6:1983.92, name path = E]{1};
	\addplot[domain=1982.6:1983.92, name path = F]{-1};      		
	\addplot[domain=1991.6:1992.25, name path = G]{1};
	\addplot[domain=1991.6:1992.25, name path = H]{-1};   	 
	\addplot[domain=2002.25:2002.92, name path = I]{1};
	\addplot[domain=2002.25:2002.92, name path = J]{-1};   			
	\addplot[domain=2008.09:2010.5, name path = K]{1};
	\addplot[domain=2008.09:2010.5, name path = L]{-1};   		
	\addplot[lightgray] fill between[of=C and D];
	\addplot[lightgray] fill between[of=E and F];
	\addplot[lightgray] fill between[of=G and H];           
	\addplot[lightgray] fill between[of=I and J];
	\addplot[lightgray] fill between[of=K and L]; 		
	\draw[dashed] (axis cs:1976,0)--(2021,0);	
	\nextgroupplot[legend to name=grouplegend1, title = B. All females]
\draw[dashed] (axis cs:1976,0)--(2021,0);
\addplot[black, very thick] table[x expr={\thisrow{year}-1}, y = selection]{results_MR_6_1.txt};
\addplot[domain=1981.0:1981.6, name path = C]{1};
\addplot[domain=1981.0:1981.6, name path = D]{-1};      
\addplot[domain=1982.6:1983.92, name path = E]{1};
\addplot[domain=1982.6:1983.92, name path = F]{-1};      		
\addplot[domain=1991.6:1992.25, name path = G]{1};
\addplot[domain=1991.6:1992.25, name path = H]{-1};   	 
\addplot[domain=2002.25:2002.92, name path = I]{1};
\addplot[domain=2002.25:2002.92, name path = J]{-1};   			
\addplot[domain=2008.09:2010.5, name path = K]{1};
\addplot[domain=2008.09:2010.5, name path = L]{-1};   		
\addplot[lightgray] fill between[of=C and D];
\addplot[lightgray] fill between[of=E and F];
\addplot[lightgray] fill between[of=G and H];           
\addplot[lightgray] fill between[of=I and J];
\addplot[lightgray] fill between[of=K and L]; 			
	\end{groupplot}
	\node at (plots c1r1.south) [inner sep=10pt,anchor=north, yshift=-5ex] {\ref{grouplegend1}};  
	\end{tikzpicture}
}
\end{center}
\par
{\small \  }
\caption{Estimation of $\protect\rho_t \protect\sigma_{E_t} $ over time.}
\label{fig:rho_sigma}
\end{figure}
}

{\small 
\begin{figure}[tbp]
\begin{center}
{\small 
\begin{tikzpicture}
	\begin{groupplot}
	[
	group style={%
		group size = 2 by 1,
		group name=plots
		, horizontal sep=1.25cm, 
		xlabels at=edge bottom,
		ylabels at=edge left,
		x descriptions at=edge bottom
	},
	set layers,cell picture=true,
	width=0.45\textwidth,
	height=0.45\textwidth,
	legend columns=-1,
	xlabel = Year,
	ylabel = Difference,
	xmin = 1975, 
	xmax = 2019,
	ymax = 0.5,
	ymin = -0.1,
	cycle list name=black white
	]
	\nextgroupplot[legend to name=grouplegend1]
	\draw[dashed] (axis cs:1976,0)--(2021,0);
	\addplot[black, very thick, double] table[x expr={\thisrow{year}-1}, y = dselection]{results_MR_6_2.txt};
	\addlegendentry{Total selection}
	\addplot[black, very thick] table[x expr={\thisrow{year}-1}, y = term1]{results_MR_6_2.txt};
	\addlegendentry{Term 1}
	\addplot[black, dashed, very thick] table[x expr={\thisrow{year}-1}, y = term2]{results_MR_6_2.txt};
	\addlegendentry{Term 2}	
	\addplot[black, dotted, very thick]  table[x expr={\thisrow{year}-1}, y = term3]{results_MR_6_2.txt};
	\addlegendentry{Term 3}		
	\addplot[black, dashdotted, very thick]  table[x expr={\thisrow{year}-1}, y = term4]{results_MR_6_2.txt};
	\addlegendentry{Term 4}			
	\addplot[domain=1981.0:1981.6, name path = C]{1};
	\addplot[domain=1981.0:1981.6, name path = D]{-1};      
	\addplot[domain=1982.6:1983.92, name path = E]{1};
	\addplot[domain=1982.6:1983.92, name path = F]{-1};      		
	\addplot[domain=1991.6:1992.25, name path = G]{1};
	\addplot[domain=1991.6:1992.25, name path = H]{-1};   	 
	\addplot[domain=2002.25:2002.92, name path = I]{1};
	\addplot[domain=2002.25:2002.92, name path = J]{-1};   			
	\addplot[domain=2008.09:2010.5, name path = K]{1};
	\addplot[domain=2008.09:2010.5, name path = L]{-1};   		
	\addplot[lightgray] fill between[of=C and D];
	\addplot[lightgray] fill between[of=E and F];
	\addplot[lightgray] fill between[of=G and H];           
	\addplot[lightgray] fill between[of=I and J];
	\addplot[lightgray] fill between[of=K and L]; 			
	\end{groupplot}
	\node at (plots c1r1.south) [inner sep=10pt,anchor=north, yshift=-5ex] {\ref{grouplegend1}};  
	\end{tikzpicture}
}
\end{center}
\par
{\small \  }
\caption{Selection effect using the Heckman-selection model and the 4 parts
of the selection effect are represented by \eqref{MR:selection}; total
sample is full population between 24 and 65 and selected population is all
workers working a positive number of hours.}
\label{fig:decomp_heckman_full}
\end{figure}
}

{\small 
\begin{figure}[tbp]
\begin{center}
{\small 
\begin{tikzpicture}
	\begin{groupplot}
	[
	group style={%
		group size = 2 by 1,
		group name=plots
		, horizontal sep=1.25cm, 
		xlabels at=edge bottom,
		ylabels at=edge left,
		x descriptions at=edge bottom
	},
	set layers,cell picture=true,
	width=0.45\textwidth,
	height=0.45\textwidth,
	legend columns=-1,
	xlabel = Year,
	ylabel = Difference,
	xmin = 1975, 
	xmax = 2020,
	ymax = 0.35,
	ymin = -0.1,
	cycle list name=black white
	]
	\nextgroupplot[legend to name=grouplegend1]
	\addplot[black, very thick, dotted] table[x expr={\thisrow{year}-1}, y = selection1]{results_MR_6_2.txt};
	\addlegendentry{Selection}
	\addplot[black, dashed, very thick] table[x expr={\thisrow{year}-1}, y = structural]{results_MR_6_2.txt};
	\addlegendentry{Structural}	
	\addplot[black, dashdotted, very thick]  table[x expr={\thisrow{year}-1}, y = composition]{results_MR_6_2.txt};
	\addlegendentry{Composition}		
	\addplot[black, very thick]  table[x expr={\thisrow{year}-1}, y = total]{results_MR_6_2.txt};
	\addlegendentry{Total}			
	\addplot[domain=1981.0:1981.6, name path = C]{1};
	\addplot[domain=1981.0:1981.6, name path = D]{-1};      
	\addplot[domain=1982.6:1983.92, name path = E]{1};
	\addplot[domain=1982.6:1983.92, name path = F]{-1};      		
	\addplot[domain=1991.6:1992.25, name path = G]{1};
	\addplot[domain=1991.6:1992.25, name path = H]{-1};   	 
	\addplot[domain=2002.25:2002.92, name path = I]{1};
	\addplot[domain=2002.25:2002.92, name path = J]{-1};   			
	\addplot[domain=2008.09:2010.5, name path = K]{1};
	\addplot[domain=2008.09:2010.5, name path = L]{-1};   		
	\addplot[lightgray] fill between[of=C and D];
	\addplot[lightgray] fill between[of=E and F];
	\addplot[lightgray] fill between[of=G and H];           
	\addplot[lightgray] fill between[of=I and J];
	\addplot[lightgray] fill between[of=K and L]; 			
	\end{groupplot}
	\node at (plots c1r1.south) [inner sep=10pt,anchor=north, yshift=-5ex] {\ref{grouplegend1}};  
	\end{tikzpicture}
}
\end{center}
\par
{\small \  }
\caption{Computation of the selection, composition and structural effect
based on equation \eqref{our:selection}-\eqref{our:structural}; total sample
is full population between 24 and 65 and selected population is all workers
working a positive number of hours.}
\label{fig:decomp_heckman1_full}
\end{figure}
}

{\small 
\begin{figure}[tbp]
\begin{center}
{\small 
\begin{tikzpicture}
	\begin{groupplot}
	[
	group style={%
		group size = 2 by 2,
		group name=plots
		, horizontal sep=1.25cm, 
		xlabels at=edge bottom,
		ylabels at=edge left,
		x descriptions at=edge bottom
	},
	set layers,cell picture=true,
	width=0.45\textwidth,
	height=0.45\textwidth,
	legend columns=-1,
	xlabel = Year,
	ylabel = Average Derivative,
	ymin = 0.1,
	ymax = 0.5,
	xmin = 1976, 
	xmax = 2020,
	cycle list name=black white
	]
	\nextgroupplot[legend to name=grouplegend1, title=A. Our method]
	\addplot[black, very thick] table[x expr={\thisrow{year}-1}, y = average]{average_derivative_2021.txt};
	\addplot[domain=1981.0:1981.6, name path = C]{1};
	\addplot[domain=1981.0:1981.6, name path = D]{-1};      
	\addplot[domain=1982.6:1983.92, name path = E]{1};
	\addplot[domain=1982.6:1983.92, name path = F]{-1};      		
	\addplot[domain=1991.6:1992.25, name path = G]{1};
	\addplot[domain=1991.6:1992.25, name path = H]{-1};   	 
	\addplot[domain=2002.25:2002.92, name path = I]{1};
	\addplot[domain=2002.25:2002.92, name path = J]{-1};   			
	\addplot[domain=2008.09:2010.5, name path = K]{1};
	\addplot[domain=2008.09:2010.5, name path = L]{-1};   		
	\addplot[lightgray] fill between[of=C and D];
	\addplot[lightgray] fill between[of=E and F];
	\addplot[lightgray] fill between[of=G and H];           
	\addplot[lightgray] fill between[of=I and J];
	\addplot[lightgray] fill between[of=K and L]; 			
		\nextgroupplot[legend to name=grouplegend1, title = B. Tobit, ymax=0.2, ymin=0]
	\addplot[black, very thick] table[x expr={\thisrow{year}-1}, y = average]{average_derivative_1_2021.txt};
	\addplot[domain=1981.0:1981.6, name path = C]{1};
	\addplot[domain=1981.0:1981.6, name path = D]{-1};      
	\addplot[domain=1982.6:1983.92, name path = E]{1};
	\addplot[domain=1982.6:1983.92, name path = F]{-1};      		
	\addplot[domain=1991.6:1992.25, name path = G]{1};
	\addplot[domain=1991.6:1992.25, name path = H]{-1};   	 
	\addplot[domain=2002.25:2002.92, name path = I]{1};
	\addplot[domain=2002.25:2002.92, name path = J]{-1};   			
	\addplot[domain=2008.09:2010.5, name path = K]{1};
	\addplot[domain=2008.09:2010.5, name path = L]{-1};   		
	\addplot[lightgray] fill between[of=C and D];
	\addplot[lightgray] fill between[of=E and F];
	\addplot[lightgray] fill between[of=G and H];           
	\addplot[lightgray] fill between[of=I and J];
	\addplot[lightgray] fill between[of=K and L]; 			
		\nextgroupplot[legend to name=grouplegend1, title = C. Propensity score, ymax = 0.2, ymin=-0.5]
	\addplot[black, very thick] table[x expr={\thisrow{year}-1}, y = average]{average_derivative_newey_2021.txt};
	\draw[dashed] (axis cs:1976,0)--(2021,0);
	\addplot[domain=1981.0:1981.6, name path = C]{1};
	\addplot[domain=1981.0:1981.6, name path = D]{-1};      
	\addplot[domain=1982.6:1983.92, name path = E]{1};
	\addplot[domain=1982.6:1983.92, name path = F]{-1};      		
	\addplot[domain=1991.6:1992.25, name path = G]{1};
	\addplot[domain=1991.6:1992.25, name path = H]{-1};   	 
	\addplot[domain=2002.25:2002.92, name path = I]{1};
	\addplot[domain=2002.25:2002.92, name path = J]{-1};   			
	\addplot[domain=2008.09:2010.5, name path = K]{1};
	\addplot[domain=2008.09:2010.5, name path = L]{-1};   		
	\addplot[lightgray] fill between[of=C and D];
	\addplot[lightgray] fill between[of=E and F];
	\addplot[lightgray] fill between[of=G and H];           
	\addplot[lightgray] fill between[of=I and J];
	\addplot[lightgray] fill between[of=K and L]; 			
	\end{groupplot}
	\node at (plots c1r1.south) [inner sep=10pt,anchor=north, xshift= 3.5cm,yshift=-5ex] {\ref{grouplegend1}};  
	\end{tikzpicture}
}
\end{center}
\par
{\small \  }
\caption{Average derivative of the wage. A. is the derivative of our control
function, B. is the average derivative using derivatives of a Tobit, C. uses
a non-separable sample selection model.}
\label{fig:average_derivative_1}
\end{figure}
}

\end{document}